\documentclass[10pt,letterpaper]{article}
\usepackage{authblk}
\usepackage[letterpaper,top=0.75in,bottom=0.75in,left=0.65in,right=0.65in]{geometry}
\usepackage{natbib}
\usepackage{amsmath}
\usepackage{amssymb}
\usepackage{bm}
\usepackage{verbatim}
\usepackage{hyperref}
\usepackage{relsize}
\usepackage{xspace}
\usepackage[caption=false]{subfig}
\usepackage[section]{placeins}
\usepackage{orcidlink}
\usepackage{xcolor}
\usepackage{newtxtext,newtxmath}

\usepackage{titlesec}
\titlespacing*{\section}{0pt}{8pt plus 2pt minus 2pt}{4pt plus 2pt minus 2pt}
\titlespacing*{\subsection}{0pt}{6pt plus 2pt minus 2pt}{3pt plus 2pt minus 2pt}
\titlespacing*{\subsubsection}{0pt}{4pt plus 2pt minus 2pt}{2pt plus 2pt minus 2pt}

\titleformat*{\section}{\normalsize\bfseries}
\titleformat*{\subsection}{\small\bfseries}
\titleformat*{\subsubsection}{\small\bfseries}

\usepackage{enumitem}
\setlist{nosep,leftmargin=3em,topsep=3pt,partopsep=0pt,parsep=0pt,itemsep=2pt}

\usepackage{etoolbox}
\AtBeginDocument{%
  \setlength{\abovedisplayskip}{6pt plus 2pt minus 2pt}
  \setlength{\belowdisplayskip}{6pt plus 2pt minus 2pt}
  \setlength{\abovedisplayshortskip}{3pt plus 1pt minus 1pt}
  \setlength{\belowdisplayshortskip}{3pt plus 1pt minus 1pt}
}

\usepackage[font=small,labelfont=bf,skip=4pt]{caption}

\let\oldbibliography\thebibliography
\renewcommand{\thebibliography}[1]{%
  \oldbibliography{#1}%
  \setlength{\itemsep}{0pt}%
  \setlength{\parskip}{0pt}%
  \small
}

\usepackage{fancyhdr}
\fancypagestyle{plain}{%
  \fancyhf{}  
  \fancyfoot[R]{LA-UR-24-33097}  
  \fancyfoot[C]{\thepage}

}

\newcommand{\vel}{\ensuremath{\mathbf{v}}\xspace}
\newcommand{\U}{\ensuremath{\mathbf{U}}\xspace}
\newcommand{\F}{\ensuremath{\mathbf{F}}\xspace}
\newcommand{\G}{\ensuremath{\mathbf{G}}\xspace}
\newcommand{\B}{\ensuremath{\mathbf{B}}\xspace}
\newcommand{\E}{\ensuremath{\mathbf{E}}\xspace}
\newcommand{\J}{\ensuremath{\mathbf{J}}\xspace}
\newcommand{\half}{\ensuremath{\frac{1}{2}}}

\newcommand{\deltax}{\ensuremath{\Delta x}}
\newcommand{\deltay}{\ensuremath{\Delta y}}
\newcommand{\deltaz}{\ensuremath{\Delta z}}
\newcommand{\deltaR}{\ensuremath{\Delta R}}
\newcommand{\iph}{\ensuremath{{i+\half}}}
\newcommand{\imh}{\ensuremath{{i-\half}}}
\newcommand{\jph}{\ensuremath{{j+\half}}}
\newcommand{\jmh}{\ensuremath{{j-\half}}}

\newcommand{\ra}{\ensuremath{\bm{r}_A}}
\newcommand{\rb}{\ensuremath{\bm{r}_B}}
\newcommand{\rp}{\ensuremath{\bm{r}}}
\newcommand{\rpp}{\ensuremath{\bm{r}{'}}}

\newcommand{\tang}{\ensuremath{\bm{t}}}
\newcommand{\norm}{\ensuremath{\bm{n}}}

\DeclareMathOperator\erf{erf}

\author[1]{Samuel W.~Jones \orcidlink{0000-0003-3970-1843}}
\affil[1]{\small Theoretical Division, Los Alamos National Laboratory,
NM, 87544, USA}
\affil[2]{\small Orchard Numerical, Vancouver, British Columbia, Canada}
\affil[3]{\small General Fusion, 6020 Russ Baker Way, Richmond, British Columbia, V7B 1B4, Canada }
\author[2]{Colin P.~M\textsuperscript{c}Nally \orcidlink{0000-0002-2565-6626}}
\author[3]{Meritt Reynolds \orcidlink{0000-0001-5880-2290}}
\title{A constrained-transport embedded boundary method for compressible resistive magnetohydrodynamics}
\date{\small Journal of Computational Physics \href{https://doi.org/10.1016/j.jcp.2025.114641}{https://doi.org/10.1016/j.jcp.2025.114641}\\Received 22 January 2025; Received in revised form 6 October 2025; Accepted 26 December 2025}

\begin{document}

\maketitle 

\begin{abstract}
    Motivated by the increased interest in pulsed-power magneto-inertial fusion devices in recent years, we present a method for implementing an arbitrarily shaped embedded boundary on a Cartesian mesh while solving the equations of compressible resistive magnetohydrodynamics. The method is built	around a finite volume formulation of the equations in which a Riemann solver is used to compute fluxes on the faces between grid cells, and a face-centered constrained transport formulation of the induction equation. The small time step problem associated with the cut cells is avoided by always computing fluxes on the faces and edges of the Cartesian mesh. We extend the method to model a moving interface between two materials with different properties using a ghost-fluid approach, and show some preliminary results including shock-wave-driven and magnetically-driven dynamical compressions of magnetohydrostatic equilibria. We present a thorough verification of the method and show that it converges at second order in the absence of discontinuities, and at first order with a discontinuity in material properties. 
\end{abstract}

\section{Introduction}
\label{sec:Introduction}

Although finite-volume constrained transport methods on Cartesian meshes have enjoyed wide success for compressible MHD, for simulating physical devices with complex geometry the scheme must handle a boundary which is not conformal the Cartesian mesh.
Examples of plasma devices with complex geometry include magnetized coaxial plasma guns \citep[also known as Marshall guns]{1960PhFl....3..134M} that form and accelerate a plasma and are used extensively for plasma generation in magnetic fusion energy devices \citep[such as those operated at General Fusion:][]{PCS16,PI3tauE}, tokamaks, stellarators, laboratory plasma experiments, and plasma etching manufacturing processes \citep{1998PhPl....5.2090S,2018PSST...27b5016S}.
Further, for applying the theory of MHD to multi-material problems with both plasma and liquid metal,
a numerical method must be capable of solving the relevant dynamical equations with parameters differing greatly between the materials.
Simulation of magneto-inertial fusion approaches such as MagLIF and Magnetized Target Fusion benefits from capabilities in this regime \citep{2016JFuE...35...69W}.
In fusion devices with liquid metal plasma facing components,
the large density contrast (up to $\sim 10^{15}$) between the plasma and liquid metal phases of the flow poses challenges for a method based on a continuous mixture between two phases, such as those applied to free-surface MHD flows 
based on the diffuse two-phase scheme of \citet{2012CS&D....5a4016D} such as those by \citet{2022Fluid...7..210S} and \citet{2023NucFu..63g6022S}.
A MHD method preserving and tracking a sharp interface separating the two phases is desirable.
Two general classes of approach to this problem are possible; either a deforming  mesh can be made conformal to the interface, or an embedded boundary method where the boundary cuts through mesh cells can be used. 
The advantage of an embedded (or immersed) boundary approach lies in the ability to employ static structured meshes for the fluids. 
In turn, this allows a separation of concerns between the meshing and resolution and interface tracking. This makes efficient implementation of the resulting method simpler.

In this work we pursue a method for compressible MHD within a finite-volume framework 
with the novelty of solving the induction equation with a constrained transport method including an embedded boundary.
For the induction equation, the solution in terms of the magnetic field is obtained using a constrained-transport scheme to locally conserve magnetic flux and guarantee divergence-free magnetic fields by construction and avoid any extra complications of ameliorating the effects of finite divergence and divergence-cleaning of the magnetic field. This is also true at material interfaces but at embedded boundaries, magnetic flux is not conserved. However, monopoles are not introduced, as we explain in Section~\ref{sec:mhd_extension}.

The method proposed in this work applies a staggered-mesh constrained-transport method for the induction equation at an embedded boundary on a Cartesian mesh. Staggered-mesh constrained transport is a proven, accurate and efficient method for MHD that has and continues to enjoy many successes in recent years \citep[][and references therein]{stone2009a,stone2024a}.
This is built on top of a cut-cell method for an embedded boundary.
There are a number of such schemes in the literature.
Noh made one of the first approaches in this area with a mix of cell-merging and flux redistribution \citep{osti_4621975}.
\citet{chenandcolella87} proposed flux-redistribution schemes to gain stability and conservation at the faces of the partial cells, and a related technique accomplishes the same by redistributing the underlying interface states \citep{2021JCoPh.42809820B}.
Cell-merging was introduced by \citep{doi:10.2514/3.9273} as a way to alleviate the small time step problem associated with cut cells by creating larger cells instead.
In this work we apply the method proposed by 
\citet{forrer1998a}.
Notably the simplicity and stability of their method arises from not treating the cut boundary cells as smaller than their full volume, at the cost of introducing local conservation errors in these cells.
However, the choice of the method for updating the continuity, momentum, and energy equations in this work is significantly independent from the choice of the method of updating the induction equation.
The \citet{forrer1998a} scheme itself has been applied before for solving compressible MHD and embedded boundaries with a cell-centered magnetic field and projection method for divergence-cleaning instead of divergence-free constrained transport \citep{2009PhPl...16h2507G,2017PhPl...24b2501L,2018PhPl...25l2513L}.
In total,  we aim to build a method based on a combination of a single-phase cut-cell moving embedded boundary method, an interface tracking method, and a ghost-fluid method for coupling the phases together. 
In this work, we present the cut-cell method for embedded boundaries with hyperbolic and parabolic physical operators in two dimensions, and extend it to moving one-and two-sided interface problems, and two-material problems.

Several approaches for solving MHD equations with a material interface have been described before.
With an inductionless approximation for low magnetic Reynolds number flow, \citet{2007JCoPh.226.1532S}   employed interface tracking, and  \citet{2014JCoPh.270..345Z,alsalamiphd} solved two-phase flows with a diffuse interface approach.
This approximation is commonly used in the study of gas bubbles in metal casting processes, for example with the interface-reconstruction volume-of-fluid methods employed by \citet{2016PhFl...28i3301J}, \citet{2022Fluid...7..349C} and others. 
However, these works avoid solving the induction equation.
An approach based on an embedded boundary condition at the material interface, as presented here, has the advantage over diffuse interface methods in that a mixed-cell closure model (which can be costly and sometimes ill-defined) is not explicitly required \citep[e.g.][]{1986IJMF...12..861B,2002JCoPh.181..577A}.

The remainder of this paper proceeds as follows:
We present the governing equations and base finite-volume method for MHD employed in Section~\ref{sec:framework} and
describe the methods for representing the embedded boundary in Section~\ref{sec:boundary}.
We review the technique for adding a embedded boundary to the finite volume solution of a system of hyperbolic conservation laws (without constraints) in Section~\ref{sec:embeddedbc}.
In Section~\ref{sec:mhd_extension} we present the technique used to implement an embedded boundary for the MHD induction equations with constrained transport.
The details of representations of the interface are discussed in Section~\ref{sec:boundary}, 
and entire the algorithm is summarized in Section~\ref{sec:algo}.
Results for a range of one and two material test problems are presented in Section~\ref{sec:results}.
Finally, Section~\ref{sec:conclusions} summarizes results and discusses further possible extensions.

\section{Basic scheme}
\label{sec:framework}

We wish to solve the equations of resistive MHD in a finite-volume framework and apply boundary conditions both in the conventional way at the edges for the mesh, and  with boundary conditions enforced at embedded, possibly moving, boundaries.
For multi-material flows, the embedded moving boundary is also a material interface, but this is limited to an issue of enforcing a boundary condition as the solution of the MHD equations in each domain is separable from the other.
In this section we present first the time semi-discretization of the scheme for multi-material flows, and then the details of the single material MHD scheme used in this work.

\subsection{Time discretization}
\label{sec:time_discretization}

We consider a material governed by the MHD equations occupying a domain bounded by an embedded boundary.
This boundary can be static or moving.
It can also have a complementary sub domain occupied by a second material governed by MHD, and in this case the motion of the interface may be dictated by the velocity of the materials.
To describe this, the required notion is for a
domain $D_\ell^t$ occupied by material $\ell$ at time $t$, and the flow state on that domain $\U|_{D_\ell^t}$.
The method globally updates all quantities on a single timestep $\Delta t$ so that $t^{n+1} = t^n + \Delta t$.
The complete update of the state $\U^n|_{D_\ell^n}$ to the next time level $\U^{n+1}|_{D_\ell^{n+1}}$ is split into two parts; the time update of the flow variables ($\U_\ell$), and, if the embedded boundary is moving, the update of the subdomain occupied by the material ($D_\ell$):
\begin{align}
\U^n|_{D_\ell^n} &\rightarrow \U^{n+1}|_{D_\ell^n} \label{eq:mhdupdate}\\
\U^{n+1}|_{D_\ell^{n}} &\rightarrow \U^{n+1}|_{D_\ell^{n+1}}\ \label{eq:interfaceupdate}.
\end{align}
The first part requires time integration of the MHD equations. 
The details of the second part vary on the type of boundary and boundary evolution considered in the problem.

The time discretization of the MHD equations in the first part of the split update equation~(\ref{eq:mhdupdate}) is handled using a method-of-lines approach with a second order strong stability preserving (SSP) Runge-Kutta (RK) time integration.  This is implemented using a low storage algorithm \citep{Gottlieb2009}.
To update the state (\U) of the MHD equations for each material from  time $t^n$ to time $t^n+\Delta t$ using a single global time step $\Delta t$ the update is:
\begin{align}
	&\U^{(1)} =  \U^n + \Delta t
	\left(\partial_t\U^n \right) \label{eq:rkstage1}\\
	&\U^{n+1} = \frac{1}{2} \U^n + \frac{1}{2} \U^{(1)} + \frac{1}{2}\Delta t
	\left(\partial_t\U^{(1)}\right) \label{eq:rkstage2} 
\end{align}
known as SSPRK(2,2) or Heun's method.
Each RK stage is computed for all domains $D_\ell$ before the next stage is computed.
The domain for material $\ell$ is constant during the update from  time $t^n$ to time $t^n+\Delta t$. 

In the second part of the split update (\ref{eq:interfaceupdate}), the interface defining the domain $D_\ell$ must be moved to the next time level and $\U_\ell$ must be defined on the new domain $D_\ell^{n+1}$.
This is trivial when the boundary is static or has a specified evolution in time.
When a level set is used to evolve a material interface at which the embedded boundary conditions are enforced the level set field $\phi$ must be time integrated to update $\phi^n$ to $\phi^{n+1}$.
In this case, the same RK scheme equations~(\ref{eq:rkstage1})--(\ref{eq:rkstage2}) using the intermediate state $\U^{(1)}$ already calculated in the MHD update equation~(\ref{eq:mhdupdate}) is used:
\begin{align}
	&\phi^{(1)} =  \phi^n + \Delta t
	\left(\partial_t \phi\left(\U^n\right) \right) \label{eq:phirkstage1}\\
	&\phi^{n+1} = \frac{1}{2} \phi^n + \frac{1}{2} \phi^{(1)} + \frac{1}{2}\Delta t
	\left(\partial_t\phi\left(\U^{(1)}\right)\right) \label{eq:phirkstage2} 
\end{align}
where $\U$ is the set of $\U_\ell$ merged into a single field (discussed further in Section~\ref{sec:levelset}). 
The restriction of $\U$ from the domain $D_\ell^{n}$ to $D_\ell^{n+1}$ is trivial for cases with a material in a moving embedded boundary, as we only consider nested domains, such that $D_\ell^{n+1} \subset D_\ell^{n}$.
For cases where the embedded boundary is a two-sided material interface, the ghost fluid method (to be discussed in Section~\ref{sec:ghost-fluid} and \ref{sec:ghost-fluid-mhd}) dictates the state of any new cut-cells in $D_\ell^{n+1}$ (i.e.\ cells in the complement $ \complement_{D_\ell^{n+1}} D_\ell^{n} $).

Thus, the method reduces in essence to what are standard methods for the MHD equations, with the addition of methods for handling the embedded boundary condition when calculating the time derivative $\partial_t\U_\ell$ terms for each material.
For most of the ensuing presentation we will omit the subscript $\ell$ and let it be understood that the given expression applies to any of the materials individually, and the subscript will be reintroduced where needed to discuss the multimaterial algorithm in.

\subsection{Spatial discretization of the MHD equations}
\label{sec:mhdspatialdisc}
The equations to be solved for resistive MHD are
\citep[see, e.g.][]{stone2009a}:
\begin{eqnarray}
	\label{eq:mass}&\partial_t\rho &+ \nabla\cdot(\rho\vel) = 0 \\
	\label{eq:momentum}&\partial_t(\rho \vel) &+ \nabla\cdot(\rho\vel\vel-\B\B+P_\mathrm{total}) = 0 \\
	\label{eq:energy}&\partial_t e &+ \nabla\cdot\left[(e+P_{\rm total})\vel-\B(\B\cdot\vel)\right] = 0 \\
	\label{eq:induction}&\partial_t\B &- \nabla\times(\vel\times\B - \eta\J) = 0 ,
\end{eqnarray}
where $P_\mathrm{total}=P_{\rm gas}+P_{\rm mag}$ is the total pressure, $P_{\rm gas}=\epsilon(\gamma-1)$ is the thermal gas pressure, $\epsilon$ the internal energy density, $\gamma$ is the ratio of specific heats, $P_{\rm mag} = \half(\B\cdot\B)$ is the magnetic pressure, $e=\epsilon + \half\rho(\vel\cdot\vel) + \half(\B\cdot\B)$ is the total energy density, $\eta$ is the resistivity expressed here as a diffusivity, and $\J=\nabla\times \B$~the current density.
Note these equations are in the conventional MHD code units, but equation~(\ref{eq:induction}) with the definition of $\J$ substituted has the same form in these units and CGS Gaussian units.
Only when the boundary motion is a function of the flow will this need to be extended to include integration of a level set field (see Section~\ref{sec:levelset}).
Because of this, the discretization of the MHD equations follows nearly unchanged from what are conventional and well established methods, and here we have implemented the scheme from \cite{stone2009a}.

We consider first ideal MHD in which the resistivity $\eta=0$. 
In flux form, the equations in two spatial dimensions are
\begin{equation}
	\partial_t\U = \dfrac{\partial}{\partial x}\F +
	\dfrac{\partial}{\partial y}\G,
	\label{eq:flux_form}
\end{equation}
where \U, \F~and \G~are
\begin{eqnarray}
	\U = \begin{pmatrix}
           \rho \\
           \rho v_x \\
           \rho v_y \\
           \rho v_z \\
	   e \\
	   B_x \\
	   B_y \\
	   B_z
         \end{pmatrix};\quad
	\F = \begin{pmatrix}
       \rho v_x \\
       \rho v_x^2 + P_{\rm total} - B_x^2 \\
       \rho v_x v_y - B_x B_y \\
       \rho v_x v_z - B_x B_z \\
	   (e+P_{\rm total})v_x - (\B\cdot\vel)B_x \\
	   0 \\
	   B_y v_x - B_x v_y \\
	   B_z v_x - B_x v_z
         \end{pmatrix};\quad
	\G = \begin{pmatrix}
           \rho v_y \\
	   \rho v_x v_y - B_y B_x \\
		\rho v_y^2 + P_{\rm total} - B_y^2 \\
           \rho v_y v_z - B_y B_z \\
	   (e+P_{\rm total})v_y - (\B\cdot\vel)B_y \\
	   B_x v_y - B_y v_x \\
	   0 \\
	   B_z v_y - B_y v_z
         \end{pmatrix}.
	 \label{eq:fluxes}
\end{eqnarray}
\\
\\
Discretizing the divergence operation and applying the divergence theorem, the
system for a uniform Cartesian mesh becomes
\begin{equation}
	\partial_t\U_{i,j} = \dfrac{\deltay}{V_{i,j}}\left(\F_{\imh,j} - \F_{\iph,j}\right) +
	\dfrac{\deltax}{V_{i,j}}\left(\G_{i,\jmh} - \G_{i,\jph}\right) + \mathcal{O}({\Delta x}^2)
    + \mathcal{O}({\Delta y}^2),
	\label{eq:flux_update}
\end{equation}
where $V_{i,j}$ is the cell volume.
We note that in the staggered-mesh constrained transport algorithm that we employ the magnetic fields are represented discretely on
the mesh as face area-averaged quantities, while $\rho$, $\vel$, $P_\mathrm{gas}$ and $e$
are represented as cell volume-averaged quantities. For completeness, we also
note that the components of the electric field \E and current density \J are represented as cell
edge-averaged quantities. This is illustrated in 
\citet{stone2009a}. 
Cell average quantities are reconstructed to faces in a piecewise linear manner following \citet{stone2009a} limiting slopes with the \cite{mignone2014a} modified van~Leer limiter.
For approximating the hydrodynamic fluxes and the intermediate state between waves we have implemented both HLL \citep[see][and references therein]{toro1997a} and HLLD \citep{2005JCoPh.208..315M} Riemann solvers in the scheme presented here with success.
We have used the constrained transport formulation described in \citet{stone2009a}, and repeat some details here as they are directly relevant in the implementation of the embedded boundary conditions.

In our algorithm, $\partial_t\U$ is given by equation~(\ref{eq:flux_update}) for
the cell-averaged variables, where \F~and \G~are obtained by solving Riemann
problems at the cell faces as in a regular Godunov code. For the face-averaged components of the state vector
(the magnetic field), we employ a constrained transport formulation, where $\partial_t\B$ is given by
\begin{equation}
	\dfrac{1}{A} \partial_t \int \B\cdot{d}\bm{s} = -\dfrac{1}{A}\int
	(\nabla\times\E)\cdot{d}\bm{s} = -\dfrac{1}{A} \oint
	\E\cdot{d}\bm{\ell},
\end{equation}
in which Stokes's theorem has been used to transform the surface integral into a line integral, and where $A$ is the area of the face where $d \bm{s}$ is the differential area, \E is an electric field (with the ideal MHD $\E=-\vel\times\B$) \footnote{Some authors when describing constrained transport MHD schemes, such as \cite{stone2009a}, invoke only an electromotive force, mathematically equivalent to this \E.}, and $d\bm{\ell}$ is the element of the line bounding the face. The discretized form for a uniform two-dimensional Cartesian mesh is
\begin{eqnarray}
	\label{eq:discreteinductionx}
	\partial_t B_{x,\imh,j} &=& -\dfrac{1}{\deltay\deltaz}
	\left(E_{z,\imh,\jph}\deltaz - E_{z,\imh,\jmh}\deltaz \right) \nonumber \\
	&=& \dfrac{1}{\deltay}\left(E_{z,\imh,\jmh} -
	E_{z,\imh,\jph}\right)
\end{eqnarray}
\begin{eqnarray}
	\label{eq:discreteinductiony}
	\partial_t B_{y,i,\jmh} &=& -\dfrac{1}{\deltax\deltaz}
	\left(E_{z,\imh,\jph}\deltaz - E_{z,\iph,\jph}\deltaz \right) \nonumber \\
	&=& \dfrac{1}{\deltax}\left(E_{z,\iph,\jph} -
	E_{z,\imh,\jph}\right) 
\end{eqnarray}
\begin{eqnarray}
	\label{eq:discreteinductionz}
	\partial_t B_{z,i,j} &=& -\dfrac{1}{\deltax\deltay}
	\left(E_{y,\iph,j}\deltay - E_{y,\imh,j}\deltay + E_{x,i,\jph}\deltax - E_{x,i,\jmh}\deltax \right)\nonumber \\
    &=& \dfrac{1}{\deltax}\left(E_{y,\imh,j} - E_{y,\iph,j}\right) + \dfrac{1}{\deltay}\left( E_{x,i,\jmh}  - E_{x,i,\jph} \right) \, .
\end{eqnarray}
The formulation of the induction equation in terms of a discrete curl of an electric field is the essence of a constrained transport method and ensures the discrete increments to $\B$ preserve the divergence-free character of an initially divergence-free field.
We follow the method of \citet{stone2009a} for calculating $E$ at cell edges in an upwinded manner which results in a stable scheme.\footnote{Note that equation~(22) of \citet{stone2009a} contains a typographical error, and the second and third (dissipation) terms should have a negative sign.}

Generalizing from the $\eta=0 $ limit, for resistive MHD we use electric field $\E=-\vel\times\B+\eta\J$, where
the current density \J~is computed by
\begin{eqnarray}
    \label{eq:jx}
	J_{x,i,\jmh} &=& \dfrac{1}{\deltay}\left( B_{z,i,j} -
	B_{z,i,j-1}\right) \\
    \label{eq:jy}
	J_{y,\imh,j} &=& \dfrac{1}{\deltax}\left( B_{z,i-1,j} -
	B_{z,i,j}\right) \\
    \label{eq:jz}
	J_{z,\imh,\jmh} &=& \dfrac{1}{\deltax\deltay}\left( \deltax B_{x,\imh,j-1} -
	\deltax B_{x,\imh,j} \nonumber \right. \\
    &\quad &\qquad\left. + \deltay B_{y,i,\jmh} - \deltay
	B_{y,i-1,\jmh}\right)~.
\end{eqnarray}
We assume that $\eta$ is a constant in both time and space, but can have different values in each material if more than one is present. It is straightforward to generalize this to a spatially-varying resistivity.
With components of $\J$ on cell edges we compute the components of the resistive electric field $\eta \J$ collocated with the inductive electric field $-\vel \times \B$ and compute the contribution to the time derivative of the magnetic field from the same discrete form of the induction equation~(\ref{eq:discreteinductionx})--(\ref{eq:discreteinductionz}).

The global time step (i.e.~the time step for the coupled ideal and resistive MHD
equations) is chosen to be
\begin{equation}
\label{eq:cfl}
	\Delta t = \left( \dfrac{1}{\Delta t_\mathrm{ideal}} + \dfrac{1}{\Delta
	t_\mathrm{res}} \right)^{-1}
\end{equation}
where
\begin{equation}
	\Delta t_\mathrm{res} = \dfrac{1}{4}\dfrac{{\Delta x}^2}{\eta}
\end{equation}
is the CFL-limited resistive time step size and $\Delta t_\mathrm{ideal}$ is the CFL-limited time step for the ideal MHD
equations
\begin{equation}
	\Delta t_\mathrm{ideal} = \dfrac{\mathrm{CFL}}{n}\min\left(\dfrac{\Delta
	x}{\max(|\bm{\lambda}_x|)}, \dfrac{\Delta y}{\max(|\bm{\lambda}_y|)}\right),
\end{equation}
where $n=2$ is the number of dimensions and $\bm{\lambda}_x$ and
$\bm{\lambda}_y$ are the eigenvalues of the ideal MHD system in the $x$- and
$y$-directions. The maximum is taken locally over the eigenvalues and minimum is taken over the whole domain, including multiple phases if they are present in the problem.

This completes the description of the time integration of the MHD equations, either in a single material, or two-material flow. 
Although we have restricted these to two dimensions for the purposes of this work, we note the elements here are not novel and have previously been documented and demonstrated in three dimensions \citep{stone2009a}.
What remains is to describe how in each RK stage of equations~(\ref{eq:rkstage1}) and~(\ref{eq:rkstage2}) the boundary conditions for the MHD state $\U$ can be computed, and how an (optional) level set field is implemented.
\section{Embedded boundary conditions - Hydrodynamics}
\label{sec:embeddedbc}
In Section~\ref{sec:framework}, we have considered the equations of MHD flow for a single, continuous medium on a Cartesian mesh with some domain boundary conditions.
In this section we describe the cut-cell method for solving (\ref{eq:flux_update}) on a Cartesian mesh with embedded boundaries in the absence of magnetic fields.
We will take as given in this section that the position of the embedded boundary has been reduced to a series of line segments cutting through cells, and that the velocity of these are known.
This allows the discussion of specific established methods for tracking the embedded interface to be deferred until Section~\ref{sec:boundary} where we describe polyline and level set implementations.
One of the biggest concerns with immersed boundary methods is that we can create grid cells with arbitrarily small \deltax~or~\deltay, which will result in a corresponding decrease in the maximum time step size (equation~\ref{eq:cfl}) one can take while maintaining numerical stability. In the present work, we have adopted the cut-cell method by \cite{forrer1998a} which avoids this so-called small cell problem because the conservative fluxes are still computed at the Cartesian cell faces, and the time integration is done for the volume averaged state of the entire Cartesian cell volume for the regular and boundary cells. In fact, the reconstruct--solve--average Godunov-like structure of the algorithm remains unchanged with the exception of an additional step to set the volume-averaged state vector for the empty (ghost) cells according to some boundary condition before the reconstruction step.

\subsection{Single--sided problems with static and moving embedded boundary}
\label{sec:single-sided}
In order to inform the state in the ghost cell, one typically requires knowledge of the state in a cell in the active part of the domain that is symmetrical to the ghost cell in the boundary.
We define the transformation $\rp\rightarrow\rpp$ as the reflection of \rp~in
the segment of the polyline closest to \rp. If this segment is bounded by nodes
$A$ and $B$ with position vectors \ra{} and \rb, then
\begin{eqnarray}
	\label{eq:reflect_point}
	&q &= \dfrac{\tang_y(r_{Bx}-r_{Ax}) -
	\tang_x(r_{By}-r_{Ay})}{(\tang_x n_y - \tang_y n_x)} \\
	&\rpp &= \rp  + 2q\norm.
\end{eqnarray}
Constraints or operations will often be most natural in either the tangential
or normal direction relative to the immersed boundary or interface. We will often want to
reflect a point \rp~in the boundary to obtain the position vector of its image
\rpp, and the state vector at the location of the image, $\U(\rpp)$, to some
order.
With an embedded boundary, the vertices of the ghost cell can be reflected in the boundary using equation~(\ref{eq:reflect_point}) into the active domain, defining what will refer to in this work as a {\sl virtual cell}.
The natural way to find values for this virtual cell would be to then
 integrate the state from the overlapping cells over the virtual cell volume.
However, a bilinear interpolation of the cell average states to the centroid of the virtual cell will be accurate to second
order and was also employed by \citet{forrer1998a}. 
The interpolation of a discrete, cell-averaged scalar field $q(x,y)$ to the point $\rpp=(x',y')$ is performed by first identifying the cell $(i',j')$ that contains the point $r'$, and the cell $(i,j)$ that will serve as the lower left cell in the interpolation stencil:
\begin{align}
    i' &= \mathrm{floor}(x'/\Delta x) + 1, \label{eq:cell-containing-i} \\
    j' &= \mathrm{floor}(y'/\Delta y) + 1, \label{eq:cell-containing-j} \\
    i &= \mathrm{nint}(x'/\Delta x), \\
    j &= \mathrm{nint}(y'/\Delta y).
\end{align}
The interpolation weights $w$ are given by
\begin{align}
    w_{11} &= \dfrac{(x_{i+1}-x')(y_{j+1}-y')}{(x_{i+1}-x_i)(y_{j+1}-y_j)}, \quad
    w_{12} = \dfrac{(x_{i+1}-x')(y'-y_{j})}{(x_{i+1}-x_i)(y_{j+1}-y_j)}, \\
    w_{21} &= \dfrac{(x'-x_{i})(y_{j+1}-y')}{(x_{i+1}-x_i)(y_{j+1}-y_j)}, \quad
    w_{22} = \dfrac{(x'-x_{i})(y'-y_{j})}{(x_{i+1}-x_i)(y_{j+1}-y_j)},
\end{align}
The interpolated value $q^*$ is then computed from the weights and the stencil as
\begin{align}
    q^* &= w_{11} q_{i,j} + w_{21} q_{i+1,j} + 
    w_{12} q_{i,j+1} + w_{22} q_{i+1,j+1}
    + \mathcal{O}(\Delta x^2)
    + \mathcal{O}(\Delta y^2),
\end{align}
and limited by a function $\sigma_{i,j}$, which is 0 if any cell in the stencil is an empty cell (therefore dropping to a first order interpolation) and 1 otherwise. This gives the final, limited value as
\begin{align}
    q' &= \sigma_{i,j}q^* + (1 - \sigma_{i,j})q_{i',j'}.
\end{align}
The state vector in a narrow band of ghost cells along the interface covering the domain of influence for the time step is set thusly and according to the particular choice of boundary condition.
Typically, the types of boundary conditions that we want to use are some
combination of Dirichlet and Neumann for the physical state variables.
For some vector fields, the local curvature of the boundary should influence the
calculation of the boundary condition and, hence, the state to be used in the
adjacent ghost cells, faces or edges.
For a ghost cell with centroid \rp{} (with its image at \rpp) a Dirichlet
condition for $\U(\bm{r}_{\rm b})$ (the state vector at the point of intersection of
($\rpp-\rp$) and the boundary) can be implemented to second order by setting
\begin{equation}
	\label{eq:dirichlet}
	\U(\rp) = 2\U(\bm{r}_{\rm b}) - \U(\rpp)\, .
\end{equation}

A boundary condition with important curvature terms arises if angular
velocity tangential to the interface is be preserved (rigid body rotation, no slip, stress-free). Then,
if the local radius of curvature of the interface is R, and the normal distance
to the interface of the cell centroid is $\deltaR$,
\begin{equation*}
	\dfrac{\vel(\rpp)\cdot\tang}{R-\deltaR} = \dfrac{\vel(\bm{r}_{\rm b})\cdot\tang}{R} =
	\dfrac{\vel(\rp)\cdot\tang}{R+\deltaR}\, ,
\end{equation*}
where $\tang$ is the surface tangent unit vector.
The ghost cell's tangential velocity would then be set using
\begin{equation}
\label{eq:solidbodydirichlet}
	\U(\rp) = (R+\deltaR)\left[2\dfrac{\U(\bm{r}_{\rm b})}{R} -
	\dfrac{\U(\rpp)}{R-\deltaR}\right].
\end{equation}
Neumann boundary conditions, where the gradient of the state vector in the
boundary-normal direction is specified at the point of intersection of
$\rpp-\rp$ and the boundary, can be generally implemented by setting
\begin{equation}
	\label{eq:neumann_curvature}
	\U(\rp) = \U(\rpp) + ((\rp-\rpp)\cdot\norm)\dfrac{\partial
	\U}{\partial(\bm{r}\cdot\norm)}\, ,
\end{equation}
where $\norm$ is the surface normal unit vector.
These methods are sufficient for implementing an embedded boundary in the mesh with a variety of boundary conditions for the compressible hydrodynamic equations.

In the case of a moving embedded boundary, where we are still interested in the dynamics on only one side of the boundary, we limit ourselves to assuming the velocity of all of the nodes of the line segments defining the boundary are specified externally to the simulation, whether that be zero, a constant or time-varying. 
That is, we do not treat a true free-surface one-sided boundary condition.
The time advance of the position of the boundary is coupled to the resistive MHD by operator splitting; i.e.~the boundary location is updated after the complete Runge-Kutta integration of the MHD equations has been performed. Every time that the boundary is moved, the cut cell volumes and centroids, cut face areas and centroids, cut edge lengths and centroids and cell types must be re-calculated.
While using a polyline (Section~\ref{sec:polyline}) renders the cost of this operation to be roughly $\mathcal{O}(n)$, it may still be a significant computational overhead depending on implementation.
Because our approach (that of \citealp{forrer1998a}) maintains the underlying
mesh's procedure for calculating fluxes on cell faces, the modifications
required to enable the boundary to move are fairly minimal. Only the boundary
conditions along the interface need to be changed. Clearly, the boundary
condition for velocity must come from the specified boundary velocity. 

\subsection{Two-sided material interface boundary}
\label{sec:ghost-fluid}
In the following and throughout the remainder of the manuscript, we refer to the embedded boundary in this scenario frequently as the (material) \emph{interface}.
For two-material problems with a moving interface, we use a ghost fluid approach \citep{fedkiw1999a} to set the boundary conditions for the solution of the MHD equations for each material and couple the state on either side of the interface. For unmagnetized flows, we solve equations (\ref{eq:mass}--\ref{eq:energy}) for each material separately with $\B=0$ using the base method described in Section~\ref{sec:framework}. We will refer to the material for which we are solving the equations for the \emph{active} material and the other material the \emph{inactive} material for the purposes of this discussion, but we note that we do solve the equations for both materials (1 and 2) during the course of the time step.
Each material requires that its ghost fluid (the state in the empty cells across the interface) be populated with a physically meaningful state vector that will result in the correct response of the active material to the presence of the inactive material.

The state vector in the ghost fluid of each material is set by considering the jump conditions at a contact discontinuity. The pressure and interface-normal velocity should be continuous, so these are simply inherited from the state in the same cell of the inactive material. We assume a no-slip condition between materials and therefore the tangential velocity should also be continuous, leading to the entire velocity vector being inherited from the inactive material. The density and energy can be discontinuous at the contact discontinuity, and the materials are permitted to have different equations of state. In this work, we limit ourselves simply to gamma-law equations of state with each material having its own adiabatic index $\gamma$. Considering now the state in material~1's ghost fluid, we perform a piecewise constant extrapolation of the entropy-like quantity
\begin{equation}
a_1 = \dfrac{P_1}{\rho_1^{\gamma_1}}
\end{equation}
into the ghost fluid cell. Next, we compute the density in the ghost fluid as 
\begin{equation}
\rho_1 = \left(\dfrac{P_2}{a_1}\right)^{1/\gamma_1}\, .
\end{equation}
The total energy density is then 
\begin{equation}
e_1 = \dfrac{P_2}{\gamma_1-1} + \half\rho_1\vel_2\cdot\vel_2\, .
\end{equation}
This completes the description of the established methods used for classes of embedded boundaries for the hydrodynamic component, and following section describes how these established methods are extended to MHD.

\section{Extension to MHD}
\label{sec:mhd_extension}
In this section we describe straightforward extensions to the method for simulating MHD flow with an immersed boundary and staggered-mesh constrained transport.
The divergences of the momentum and energy fluxes in a cell consist of terms in all three components of \B. In our finite volume Godunov approach following \citet{stone2009a}, the Riemann solver computes the fluxes across a face using $\B^*$, the magnetic field in the intermediate state between the two waves that bound the face during the time step. 
The longitudinal component of \B~at the face is taken directly from the co-located face-averaged state, while the two perpendicular components are computed from the spatial reconstruction of their cell-averaged values.
In 2D, $B_z$ can be treated as though it is effectively a cell averaged quantity because $\partial_z B_z=0$, and therefore its reconstruction is treated like that of any of the volume-averaged quantities.

\subsection{MHD with a one-sided static or moving embedded boundary}
The basic boundary condition for the  ideal MHD operator in the induction equation is one which can be described as a perfectly conducting wall. This implies that the boundary-tangential magnetic field is zero on the boundary, and only a nonzero specified boundary-normal magnetic field is present.
The velocity at the wall is also given and is the velocity of the boundary, so that magnetic field lines penetrating the boundary are pinned to the boundary material points.
The term which must be computed  in the induction equation is $\nabla\times \E =\nabla\times(-\vel\times\B)$.
This can be accomplished by setting, for the velocity, a Dirichlet condition $\vel=\vel_{\rm b}$ with $\vel_{\rm b}$ the velocity of the (possibly stationary) boundary, and for the tangential and normal components of the magnetic field two separate Dirichlet boundary conditions: $\B\cdot\tang=0$ for the boundary-tangential component and $\B\cdot\norm=B_n$ for the boundary-normal component, where $B_n$ is specified.
These can be enforced in the \citeauthor{forrer1998a} cut-cell approach in the same way as for the hydrodynamic operators by using equation~(\ref{eq:dirichlet}) or extensions thereof, such as one for preserving a current-free condition at the boundary, explained below. 
The electric field calculation and constrained-transport update for the magnetic field can then be done identically in active, inactive (ghost) and cut cells, faces and edges.

For the magnetic field, if we were to impose the boundary condition on the tangential component of the magnetic field using equation~(\ref{eq:dirichlet}), then implementing this condition to a current-free magnetic field with a  $1/R$ dependence in the domain would still result in an extrapolation to the ghost faces incurring a spurious nonzero current.
We wish to impose the Dirichlet boundary condition in a way which preserves current-free steady state solutions.
If we consider the current density, in cylindrical $(r,\phi,z)$ coordinates, in the $x$-$y$ plane, and demand it be zero we have
\begin{equation}
	\J = \nabla\times\B = -\dfrac{\partial B_z}{\partial r}\hat{\phi} +
	\dfrac{1}{r}\dfrac{\partial (r B_\phi)}{\partial r} \hat{z} = 0\,.
\end{equation}
Thus to ensure the boundary condition for the azimuthal (tangential) component can respect 
\begin{eqnarray}
    \dfrac{\partial(r B_\phi)}{\partial r} = 0\, ,
\end{eqnarray}
and discretely reproduce the steady state we replace equation~(\ref{eq:dirichlet}) with
\begin{equation}
	\label{eq:dirichlet_curvature}
	\U(\rp) = \dfrac{1}{R+\deltaR}\left[2R\U(\bm{r}_b) - (R-\deltaR)\U(\rpp)\right]\ ,
\end{equation}
for the magnetic field where again the local radius of curvature of the interface is R, and the normal distance
to the interface of the cell centroid is $\deltaR$.
This is analogous to the constant angular velocity (solid body rotation) situation in equation~(\ref{eq:solidbodydirichlet}).

By taking this approach, although the extrapolation of $\B$~will result in a non-solenoidal magnetic field locally at the boundary in the ghost cells, no such error is introduced into $\B$ on the physical part of the domain because we still calculate updates to the active mesh face B-fields by taking the discrete curl of $\E$, and thus exploiting $\nabla\cdot(\nabla\times\E)=0$ in-keeping with our underlying constrained transport approach.

Generalizing, the induction equation including ohmic resistivity $\eta$~is
\begin{equation}
	\partial_t\B - \nabla\times(\vel\times\B - \eta\J) = 0,
\end{equation}
where $\eta\J$ is the electric field from Ohm's law. We now turn our attention to computing the parabolic part of the equation $\nabla\times\eta\J = \nabla\times(\eta\nabla\times\B)$. We note again that the current density is an edge-centered variable that is computed discretely on the uniform Cartesian mesh as in (\ref{eq:jx}--\ref{eq:jz}).
The ohmic electric field, and hence \J($=\nabla\times\B$), is needed for a narrow band of ghost edges on the
inactive side of the immersed boundary so that its curl can be taken discretely, just as was described for the ideal term.

Because our formulation of constrained transport uses a staggered mesh, we need to interpolate the components of (\B) to virtual face or cell centers for certain boundary conditions.
We interpolate the $B_z$ to the virtual cell center like other cell-averaged quantities as described in Section~\ref{sec:single-sided}.
For the face-centered $B_x$, a stencil consisting of $x$-normal faces whose centers bound the virtual point $\rpp=(x',y')$ is chosen:
\begin{align}
    \imh &= \mathrm{nint}\left(\left((x'+\half\Delta x\right)/\Delta x\right), \\
    j &= \mathrm{nint}(y'/\Delta y),
\end{align}
while the cell containing $\rpp$~$(i',j')$~is defined by equations~(\ref{eq:cell-containing-i}-\ref{eq:cell-containing-j}).
The interpolation weights $w$ are given by
\begin{align}
    w_{11} &= \dfrac{(x_\iph-x')(y_{j+1}-y')}{(x_\iph-x_\imh)(y_{j+1}-y_j)}, \quad
    w_{12} = \dfrac{(x_\iph-x')(y'-y_{j})}{(x_\iph-x_\imh)(y_{j+1}-y_j)}, \\
    w_{21} &= \dfrac{(x'-x_\imh)(y_{j+1}-y')}{(x_\iph-x_\imh)(y_{j+1}-y_j)}, \quad
    w_{22} = \dfrac{(x'-x_\imh)(y'-y_{j})}{(x_\iph-x_\imh)(y_{j+1}-y_j)},
\end{align}
yielding the interpolated value
\begin{align}
    B_x^* &= w_{11} B_{x,\imh,j} + w_{21} B_{x,\iph,j} + 
    w_{12} B_{x,\imh,j+1} + w_{22} B_{x,\iph,j+1}
    + \mathcal{O}(\Delta x^2)
    + \mathcal{O}(\Delta y^2).
\end{align}
This is again limited by a function $\sigma_{\imh,j}$, which is 0 if any face in the stencil is an empty face (therefore dropping to a first order interpolation) and 1 otherwise:
\begin{align}
    B_x' &= \sigma_{\imh,j}B_x^* + (1 - \sigma_{\imh,j})B_{x,i'-\half,j'}.
\end{align}
An analogous procedure is followed to interpolate $B_y$ to $\rpp$ from the $y$-normal faces.
Once \B~is defined in the ghost cells and faces, \J~is computed at cell
edges including cells in a narrow band on the inactive side of the immersed
boundary (the ghost edges). 

This completes the description of how the boundary conditions for the ideal MHD and Ohmic contributions to the electric field are imposed separately for one-sided boundaries.
After discussion of the details of handling the representation of the interface in Section~\ref{sec:boundary}, the algorithm is summarized for clarity in full in Section~\ref{sec:polylinealgo}.

\subsection{MHD extensions to ghost fluid method for two-sided embedded boundary}
\label{sec:ghost-fluid-mhd}
For a subset of problems, we are interested in evolving the state on both sides of an embedded boundary. For example, in a plasma and a liquid metal. This situation requires an extension of the ghost fluid algorithm presented in Section~\ref{sec:ghost-fluid}. In addition to having different equations of state (or in our case, different values of $\gamma$), the materials are also permitted to have different resistivities $\eta$.
At the material interface, the electric field $-\vel\times\B+\eta\J$ may be discontinuous owing to jumps in the tangential component of $\B$~or a jump in the material resistivity $\eta$. Evolving separate $\B$~fields on either side of the interface is problematic because decomposing the magnetic field into normal and tangential components will be accurate only to truncation error, resulting in an error in $\nabla\cdot\B$ that is significantly larger than roundoff and introducing monopoles. Instead, we evolve one unified global $\B$~field but incorporate some of the discontinuous aspects by computing separate per-material edge-centered electric fields and combining them using an edge occupation fraction-weighted average.
We note that this poses the limitation that while the a sharp interface method such as the ghost fluid method will eliminate numerical dissipation of mass, the present multi-material treatment will not eliminate numerical resistivity at the interface in the same way. Such a method is desirable, however, and would be a welcome extension to the present work.
An alternative approach would be to compute volume- and area-fraction weighted averages of the velocity field and resistivity and computing a global electric field from this. However, in practice we found that computing separate electric fields and averaging them produced better results.
This algorithm is summarized for clarity in full in Section~\ref{sec:levelsetalgo}.

\section{Representation of the embedded boundary}
\label{sec:boundary}
To track the location of the embedded boundary in different problems, we employ either a polyline (supplementary Lagrangian mesh) or a level set.
Specifically, we employ a polyline representation for static or specified motion one-sided embedded boundary problems, due to its simplicity when restricted to those cases.
However, for two-sided (material interface) problems, where the interface motion maybe driven by the velocity field, we employ a level-set method to track the embedded boundary. 
Both of these methods are well developed in the literature, but we nevertheless describe the specific implementations used in this work here.
This also serves to demonstrate that the cut-cell and ghost-fluid methods employed are independent of how the identification of the boundary location is accomplished.

\subsection{Supplementary Lagrangian mesh - polyline}
\label{sec:polyline}
\begin{figure}
	\centering
	\includegraphics[width=.7\textwidth]{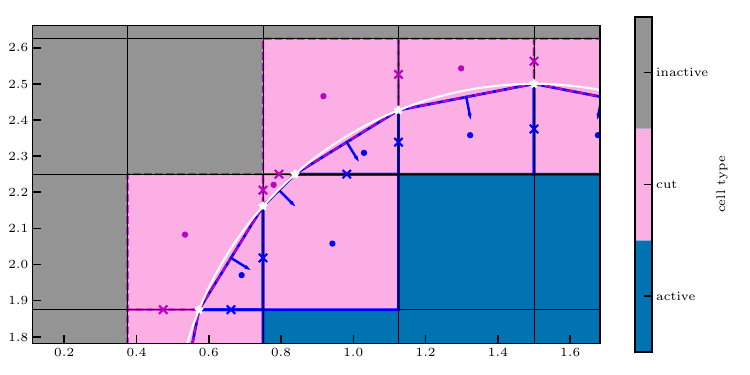}
	\caption{Illustration of the process of dividing the cells in our cut
	cell algorithm implementation. The cell type is indicated by cell color (see color map). The polyline representation of the embedded boundary is drawn in white, and white stars mark where it intersects a cell face. The outer and inner faces (lines) and centroids (points) of the cut cells are colored magenta and blue, respectively. The cut face centroids are marked with crosses in the corresponding colors. The boundary normal vectors for the discretized piecewise-linear boundary are drawn as blue arrows.}
	\label{fig:cutcell_plot}
\end{figure}
The polyline defines the boundary as a closed polygon with nodes specified in anticlockwise order, with nodes that may lie both interior to or exterior to the computational domain. The advantages of a supplementary Lagrangian representation are that it requires little storage, trivially provides the local boundary tangent and normal vectors, and for immersed interface applications where the boundary can move, the boundary is not distorted by numerical diffusion. However, the arbitrary node locations mean that care should be taken in the resolution of the Lagrangian mesh and how it relates to the resolution of the Eulerian mesh. This is particularly apparent when conducting resolution studies. One approach would be to ensure that the node spacing of the Lagrangian mesh was much finer than the width of a cell on the finest Eulerian mesh to be used. While this may appear to have little computational cost because of the lower dimensionality of the Lagrangian mesh, the expense of the method can escalate quickly because operations such as searching for the nearest node(s) to an Eulerian cell must be performed every time the Lagrangian mesh moves.

Once the polyline describing the boundary has been specified, we must identify
all of the boundary cells cut by the polyline. This is accomplished by traversing the polyline in small increments $\delta \vec{p}$ and checking whether we
have crossed a cell face. If this is the case, the cells bounding the face are
flagged as boundary cells. In fact, each increment $\delta \vec{p}$ is performed
as two steps of $(\delta \vec{p}\cdot\hat{i})\hat{i}$ and $(\delta
\vec{p}\cdot\hat{j})\hat{j}$, where $\hat{i}$ and $\hat{j}$ are unit vectors in
the $x$ and $y$ directions, respectively. This is to avoid the situation whereby
the boundary exactly cuts through the corner of a cell, which can lead to
undefined and undesirable consequences later on.
There may be cases where the polyline resolution is finer than that of the
Cartesian mesh, in which case several nodes may occupy a single grid cell. This
is handled by representing the boundary as a linear function within
each cell, connecting the entry and exit points of the cell where the polyline
cut through the edges. Each boundary cell then contains a complete local
representation of the boundary, which consists of the interface normal unit
vector and the midpoint of the interface within the cell.
For the remaining cells, their type can be determined by computing the cross
product of the direction vector of all polyline segments with the position
vector of the cell centroid relative to the tail of the polyline segment. For
example, for a segment $\vec{AB}$ connecting nodes with position vectors
\ra{} and \rb, the type of cells with centroid at position vector \rp{} are
determined as follows
\begin{equation}
	\mathrm{cell\ type} =
	\begin{cases}
		\text{empty} & \text{if}\
		(({\rb}_i-{\ra}_i)\times({\rb}_i-\rp))\cdot\hat{k} > 0\ 
		\quad \forall~1\leq i \leq N\\
		\text{regular} & \text{if}\
		(({\rb}_i-{\ra}_i)\times({\rb}_i-\rp))\cdot\hat{k} < 0\
		\quad \forall~1\leq i \leq N
	\end{cases}
\end{equation}
where $N$ is the number of nodes in the closed polyline and $\hat{k}$ is a unit
vector in the $z$-direction. We note that this will only work with closed
polylines that describe convex polygons.

In Figure~\ref{fig:cutcell_plot} we present an illustration of the process of
dividing the cells given a polyline. The color scale indicates the cell type,
with blue being regular cells, and pink being boundary cells. The gray cells are
empty and are completely immersed in the inactive boundary region. The white curve is the polyline that was used to initialize the boundary, and its discrete
representation is drawn with a dashed blue line within each boundary cell. The
centroids of the active and inactive polygons within each boundary cell is
indicated with either a blue (active) or magenta (inactive) dot. For each face
that is cut by the polyline, the inactive and active face centroids are
indicated by crosses of the corresponding color.

In this work we also use the polyline representation for the embedded boundary for problems with a specified motion of the interface, where the velocity of the boundary segment is needed for various boundary condition.
In this case, the velocity of a polyline segment is computed as the arithmetic
mean of the velocities of its bounding nodes, which is accurate to second order.

\subsection{Level set method}
\label{sec:levelset}
For problems with a moving material interface we opted to implement a level set representation because of the ease with which it can represent large  distortions in the geometry of the material interface and avoid the inevitable mesh-tangling of a Lagrangian mesh for problems involving complex flows. It also proved to be just as straightforward for static immersed boundaries. The level set method is at this point in time mature and well documented in the literature. Therefore we refer the interested reader to the comprehensive book by \citet{osher2004a} and will restrict the description in our manuscript to basic concepts as necessary and specific details of our particular implementation.

We define and store a cell-centered scalar function $\phi$ whose value is the signed distance to the material interface (referred frequently hereafter as `the interface'). On one side of the interface the function is positive and on the other it is negative. Therefore, it is trivial to evaluate whether or not a point is inside or outside of the interface, and the location of the interface is given by the zero level set of this scalar function. 
Thus, the procedure required with the polyline for identifying boundary cells is unnecessary when using the levelset.
At the cell-level, the interface is treated as planar, although curvature effects may still be included in the matching and boundary conditions as long as the local radius of curvature is calculated from the level set. The scalar field $\phi$~is interpolated to the cell edges and the zero-crossings are identified on the edges\footnote{We note that in 2D the $xy$~edges are co-located with the cell corners and the $xz$~and $yz$~edges are co-located with the y-normal and x-normal faces, respectively}. The interface can then be approximated as a line (plane in 3D) connecting these zero-crossings. This ensures continuity of the interface across neighboring cells. Any cell/face that is connected to an edge with a zero crossing of $\phi$~is labelled as a cut cell/face.
Volume fractions are computed using the resulting polygon (in 2D) areas. The interface normal vector, required for the reflection of the ghost cell in the boundary and to decompose various quantities into components tangential and normal to the interface, is calculated as $\hat{\bm{n}}= \nabla\phi/|\nabla \phi|$. This is computed in all cells using a dimensionally split modified van Leer limited reconstruction of $\phi$ \cite{mignone2014a}, which is just the van Leer slope limiter ($\varphi$) in the uniform Cartesian mesh limit:
\begin{eqnarray}
    \varphi(v) = \dfrac{v + |v|}{1 + |v|}; \quad
    v = \dfrac{U_{i+1}-U_i}{U_i - U_{i-1}}.
\end{eqnarray}

Movement of the interface is described by the level set equation
\begin{equation}
    \partial_t\phi + \nabla\cdot(\vel\phi) = \phi(\nabla\cdot\vel)\, . \label{eq:levelset}
\end{equation}
As mentioned in Section~\ref{sec:time_discretization}, at each RK stage  (equations~\ref{eq:rkstage1} and~\ref{eq:rkstage2}) of the MHD update, a global cell-averaged velocity vector is constructed as a volume fraction weighted average of the velocity in the two materials and stored for use in the time integration of equation~(\ref{eq:levelset}) with the low-storage RK scheme equation~(\ref{eq:phirkstage1}) and~(\ref{eq:phirkstage2}).
Equation (\ref{eq:levelset}) is solved using the finite volume discretization
\begin{eqnarray}
    \partial_t\phi_{i,j} = 
    \dfrac{1}{V_{i,j}}
    \left[
    -\Delta y \left(F^\phi_\iph - F^\phi_\imh\right)
    -\Delta x \left(G^\phi_\jph - G^\phi_\jmh\right) 
    \right.\nonumber\\
    \left.
    + \Delta y\left(v_{x,\iph}^{up} - v_{x,\imh}^{up}\right)
    + \Delta x\left(v_{y,\jph}^{up} - v_{y,\jmh}^{up}\right)
    \right]
\end{eqnarray}
in space. $v^{up}$ are the upwind velocities are defined by
\begin{eqnarray}
    v^{up}_{x,\imh} = 
	\begin{cases}
        v_{x,\imh}^-, & F^\phi_\imh \geq 0\\
        v_{x,\imh}^+, & F^\phi_\imh < 0
	\end{cases}, \\
    v^{up}_{y,\jmh} = 
	\begin{cases}
        v_{y,\jmh}^-, & G^\phi_\jmh \geq 0\\
        v_{y,\jmh}^+, & G^\phi_\jmh < 0
	\end{cases},
\end{eqnarray}
where superscript $\pm$ indicates the value on the lower or upper side of a face.
The fluxes $F^\phi$ and $G^\phi$ are 
approximated using a simple local Lax-Friedrichs flux:
\begin{eqnarray}
    F^\phi_\imh = \dfrac{1}{2}\left(
    v^-_{x,\imh}\phi^-_\imh + v^+_{x,\imh}\phi^+_\imh 
    - \dfrac{\Delta x}{2\Delta t}\left(
    \phi^+_\imh - \phi^-_\imh
    \right)
    \right)
\end{eqnarray}
The face-adjacent variables themselves are reconstructed and limited from a stencil of cell-averages using a $5^\mathrm{th}$-order WENO reconstruction \citep{shu2009a,borges2008a}, which we have found to be of great benefit in improving the accuracy and robustness of the simulations.

It is known that $\phi$ will deviate from a signed distance function as a result of numerical dissipation and must periodically be corrected in a procedure known as reinitialization.
We follow the approach described in \citet{sussman1999}, which is based on integrating the Eikonal equation
\begin{equation}
    \partial_\tau\phi = S(\phi_0)(1-|\nabla\phi|),
    \label{eq:Eikonal}
\end{equation}
forwards in pseudotime $\tau$, in which $S(\phi)$ is a sign function. $\phi_0$ is the level set passive scalar field at pseudotime $\tau=0$.
This is a Hamilton-Jacobi equation that propagates information away from the interface with speed 1. Since $\phi$ need only be a signed distance function in a narrow band of cells close to the interface and simply positive or negative elsewhere depending on which side of the interface a cell is on, the number of pseudotime steps needed is on the order of the number of cells required to be in this narrow band.

The implementation of $S(\phi)$ uses a Heaviside function:
\begin{eqnarray}
    S(\phi) &= 2\left(H(\phi, \Delta x) - \half\right), \\
    H(\phi, \Delta x) &= \max\left(0,
    \min\left(
    1,
    \dfrac{1}{2}\left(
    1 + \dfrac{\phi}{\Delta x} + 
    \dfrac{1}{\pi}\sin\left(\dfrac{\pi\phi}{\Delta x}
    \right)
    \right)
    \right)
    \right),
\end{eqnarray}
and we use a second order Godunov/MUSCL upwind difference approximation to the $\nabla\phi$, as is typical for discretized Hamilton-Jacobi equations \citep{sethian1999a}.
Considering first the $x$-direction, using a stencil of width 5 cells, second order backward and forward difference approximations and the left ($a$) and right ($b$) faces are calculated:
\begin{align}
    s_l^a &= U_{i,j} - 2U_{i-1,j}
    + U_{i-2,j}, \quad
    s_r^a = U_{i+1,j} - 2U_{i,j}
    + U_{i-1,j}, \\
    s_l^b &= U_{i+1,j} - 2U_{i,j}
    + U_{i-1,j}, \quad
    s_r^b = U_{i+2,j} - 2U_{i+1,j}
    + U_{i,j}.
\end{align}
and a MUSCL-style limiting approach is applied to give the slopes on the left and right faces ($a$ and $b$, respectively):
\begin{align}
    s^a &= \mathrm{minmod}(s_l^a, s_r^a), \quad
    s^b = \mathrm{minmod}(s_l^b, s_r^b), \\
    a &= U_{i,j} - U_{i-1,j} + \half s^a, \quad
    b = U_{i+1,j} - U_{i,j} - \half s^b.
\end{align}
The same procedure is applied to calculate the slopes in the $y$-direction at the bottom and top faces, $c$ and $d$, respectively.
A Godunov splitting is applied to the slopes:
\begin{align}
    a^\pm = \half(a \pm |a|),\quad
    b^\pm = \half(b \pm |b|),\quad 
    c^\pm = \half(c \pm |c|),\quad 
    d^\pm = \half(d \pm |d|).
\end{align}
The Godunov 2D rule for the Eikonal Hamiltonian ($H(\nabla\phi)=|\nabla\phi|$) gives two gradient magnitudes, one for the case where the speed $S(\phi)$ is positive:
\begin{align}
    |\nabla^+\phi| = \dfrac{1}{\Delta x}\sqrt{
    \max((a^+)^2, (b^-)^2) + \max((c^+)^2, (d^-)^2)
    },
\end{align}
(where the factor of $1/\Delta x$ comes from the fact we have a uniform stencil and were working with differences as opposed to gradients) and one for where the speed is negative:
\begin{align}
    |\nabla^-\phi| = \dfrac{1}{\Delta x}\sqrt{
    \max((a^+)^2, (b^-)^2) + \max((c^+)^2, (d^-)^2)
    }.
\end{align}
Thus, the discretized form of (\ref{eq:Eikonal}) is
\begin{align}
    \partial_\tau\phi_{i,j} = 
    \begin{cases}
        S(\phi_{0,i,j}) (1 - |\nabla^+\phi_{i,j}|), & S(\phi_{0,i,j})\geq 0, \\
        S(\phi_{0,i,j}) (1 - |\nabla^-\phi_{i,j}|), & S(\phi_{0,i,j})<0.
    \end{cases}
\end{align}
This is discretized using a method of lines approach and integrated forwards in pseudotime using the same low storage RK2 algorithm described in Section~\ref{sec:time_discretization} with a pseudo-time step of $\Delta t = 0.25 \Delta x$. 
We found that it is not practically necessary to integrate in pseudo-time until a steady-state for $\phi$ has been achieved for the reinitialization to be satisfactory.
In this work, we chose to take only 10 pseudo-time steps when re-initializing the level-set because (1) $\phi$ needs to be a signed distance function only in a narrow band ($\sim 5$ cells) around the interface, and (2) the numerical deviation of $\phi$ from being a signed distance function over the course of an advection step is usually quite minimal. When initializing $\phi$ at the beginning of a simulation, however, we take 400 pseudo-time steps to ensure that $\phi$ is well defined over a larger scale.
The value of $\phi$ in the cut cells does not change during the pseudo-time integration so that the interface moves minimally during the reinitialization procedure.

\section{Summary of Algorithms}
\label{sec:algo}

\subsection{Algorithm for MHD with a one-sided static or moving embedded boundary}
\label{sec:polylinealgo}
The algorithm for evolution is laid out below for clarity. Details of program control and output are omitted for brevity.
\begin{enumerate}
    \item[Step 1] Using the polyline, flag cells as active, cut or inactive and record the nearest polyline segment to each cell (Section~\ref{sec:polyline}).
    \item[Step 2] Initialize entire state $\U_0$ in primitive variables in active and cut cells and in faces bounding active and cut cells.
    \item[Step 3]  $\U^n|_{D_\ell^n} \rightarrow  \U^{n+1}|_{D_\ell^{n+1}}$ (Section~\ref{sec:time_discretization}) 
    \begin{enumerate}
        \item[Step 3(a)] Determine $\Delta t$ from equation~(\ref{eq:cfl})
        \item[Step 3(b)] Fill $\U_0$ state in empty cells and empty faces that do not bound cut or active cells in the ghost region (up to $5\Delta x$ away from embedded boundary) based on choice of boundary condition; recompute the cell average \B in active, cut and ghost cells (Section~\ref{sec:embeddedbc}).
        \item[Step 3(c)] Copy $\U_0\rightarrow\U_1$ and convert $\U_0$ to conserved form.
        \item[Step 3(d)] $\U^n|_{D_\ell^n} \rightarrow \U^{n+1}|_{D_\ell^n}$ (RK stages equations \ref{eq:rkstage1}--\ref{eq:rkstage2})
        \label{enum:polystageloop}
        \begin{enumerate}
            \item[Step 3(d)i]  \label{enum:setbcs} Set domain boundary conditions in $\U_1$ (if active region is adjacent to a domain boundary).
            \item[Step 3(d)ii.] Calculate fluxes and \E~ at cell face from $\U_1$ on active, cut and ghost faces, omitting last ghost face where bounding state is undefined.
            \item[Step 3(d)iii]  Calculate cell average components of $\partial_t\U_1$
            \item[Step 3(d)iv]  Calculate ideal component of \E~on cell edges of active, cut and ghost cells including upwind dissipation terms (Section~\ref{sec:mhdspatialdisc}).
            \item[Step 3(d)v] Calculate \J~and Ohmic \E~on cell edges, add to  ideal \E.
            \item[Step 3(d)vi]  Calculate face-averaged $\partial_t B_x$, $\partial_t B_y$, and cell-averaged $\partial_t B_z$.
            \item[Step 3(d)vii]  Convert $\U_1$ to conserved form for low-storage RK update.
            \item[Step 3(d)viii]  Update active and cut cell averages and face-averages on faces bounding active and cut cells in $\U_1$ to next RK stage.
            \item[Step 3(d)ix]  Set \B~on ghost faces in updated $\U_1$ (Section~\ref{sec:embeddedbc}) and update cell-averaged \B in $\U_1$.
            \item[Step 3(d)x]  Convert $\U_1$ back to primitive variables on material meshes.
            \item[Step 3(d)xi]  Fill cell-averaged state in ghost cells (Section~\ref{sec:embeddedbc})
            \item[Step 3(d)xii]  If not last RK stage, go to Step~3(d)i.
        \end{enumerate}
        \item[Step 3(e)] $\U^{n+1}|_{D_\ell^{n}} \rightarrow \U^{n+1}|_{D_\ell^{n+1}}$: Update polyline node locations and re-define cell and face types (equation~\ref{eq:interfaceupdate}).\footnote{With a specified interface motion, it is also simple to do this update inside the integration stages of Step~\ref{enum:polystageloop} to reduce splitting errors.}
        \item[Step 3(f)] Update to $t^{n+1}$ complete, Copy $\U_1\rightarrow\U_0$.
    \end{enumerate}
\end{enumerate}

\subsection{Algorithm for MHD with a two-sided static or moving embedded boundary}
\label{sec:levelsetalgo}
The algorithm for the two-sided ghost fluid MHD method is laid out below for clarity. These are largely the same steps as the previous subsection, but with the addition of handling multiple materials. There are three mesh objects $M_\ell$ each with its own state fields $\U^n|_{D_\ell^{n}}$: one mesh per material and a master mesh ($M_0$) that owns the \B~and level set fields, and has a state defined over the entire domain of all materials ($D_0 = \cup_{\ell>0} D_\ell $).
\begin{enumerate}
    \item[Step 1] On $M_0$ initialize the level set scalar field as $\pm\Delta x$ and propagate by solving the Eikonal equation (Section~\ref{sec:levelset}).
    \item[Step 2] On $M_{\ell > 0}$, using the level set field, flag cells as active, cut or inactive and record the unit normal and tangent vectors to the interface.
    \item[Step 3] \label{enum:gfm-start} On $M_{\ell > 0}$, initialize entire state $\U_0$ for each material in primitive variables in active and cut cells and in faces bounding active and cut cells. On $M_{0}$ initialize \B.
    \item[Step 4]  $\U^n|_{D_\ell^n} \rightarrow  \U^{n+1}|_{D_\ell^{n+1}}$ (Section~\ref{sec:time_discretization})
    \begin{enumerate}
        \item[Step 4(a)] Determine $\Delta t$ from equation~(\ref{eq:cfl})
        \item[Step 4(b)] On $M_{\ell > 0}$ fill $\U_0$ state in empty cells and empty faces that do not bound cut or active cells in the ghost region (up to $5\Delta x$ away from embedded boundary) based on state in other material and the jump conditions (Section~\ref{sec:ghost-fluid}); copy \B~from $M_0$.
        \item[Step 4(c)] On $M_{\ell > 0}$, copy $\U_0\rightarrow\U_1$ and convert $\U_0$ to conserved form.
        \item[Step 4(d)] $\U^n|_{D_\ell^n} \rightarrow \U^{n+1}|_{D_\ell^n}$ (RK stages equations \ref{eq:rkstage1}--\ref{eq:rkstage2})
        \begin{enumerate}
            \item[Step 4(d)i] \label{enum:gfm-setbcs} $\forall M$ set domain (mesh edge) boundary conditions
            \item[Step 4(d)ii] On $M_{\ell > 0}$, calculate fluxes and \E,~\vel at cell face from $\U_1$ on active, cut and ghost faces, omitting last ghost face where bounding state is undefined.
            \item[Step 4(d)iii] On $M_{\ell > 0}$, calculate cell average components of $\partial_t\U_1$ (Equations~\ref{eq:mass}-\ref{eq:energy})
            \item[Step 4(d)iv] On $M_{\ell > 0}$, calculate ideal \E~on cell edges of active, cut and ghost cells including upwind dissipation terms (Section~\ref{sec:mhdspatialdisc}).
            \item[Step 4(d)v] On $M_{0}$, calculate \J~on the master mesh and copy to material meshes.
            \item[Step 4(d)vi] On $M_{\ell > 0}$, calculate Ohmic contribution to \E~on cell edges for each material and add Ohmic \E~to ideal \E.
            \item[Step 4(d)vii] Combine \E~on $M_{\ell > 0}$ into a single \E~field on $M_0$ using an edge-fraction weighted sum.
            \item[Step 4(d)viii] On $M_{0}$, calculate face-averaged $\partial_t B_x$, $\partial_t B_y$, and cell-averaged $\partial_t B_z$ components of $\partial_t\U_1$ (Equation~\ref{eq:induction}) for faces bounding active or cut cells on master mesh.
            \item[Step 4(d)ix] $\forall M$, convert $\U_1$ to conserved form for low-storage RK update.
            \item[Step 4(d)x] On $M_{\ell > 0}$, update active and cut cell averages in $\U_1$ to next RK stage for all materials (Equations~\ref{eq:rkstage1} and \ref{eq:rkstage2}).
            \item[Step 4(d)xi]  On $M_0$ update all face-averaged and cell-centered ($B_z$) \B-fields to next RK stage.
            \item[Step 4(d)xii]  Copy updated \B~from $M_0$ to $M_{\ell > 0}$  material meshes and re-compute cell-averaged~\B.
            \item[Step 4(d)xiii]  On $M_{\ell > 0}$ fill $\U_1$ state in empty cells and empty faces that do not bound cut or active cells in the ghost region (up to $5\Delta x$ away from embedded boundary) based on state in other material and the jump conditions (Section~\ref{sec:ghost-fluid}).
            \item[Step 4(d)xiv]  On $M_0$ set full cell-averaged state and face-averaged $\vel$~on entire mesh by volume-fraction and area-fraction weighted sum from $M_{\ell>0}$.
            \item[Step 4(d)xv] On $M_{\ell > 0}$ convert $\U_1$ to primitive form.
            \item[Step 4(d)xvi] If not last RK stage, go to Step~4(d)i.
        \end{enumerate}
        \item[Step 4(e)] \label{enum:gfm-end} $\U^{n+1}|_{D_\ell^{n}} \rightarrow \U^{n+1}|_{D_\ell^{n+1}}$: On $M_0$ perform RK time integration of level set scalar field advection using face averaged~\vel for Riemann solution fluxes. Reinitialize level set if needed. (solve equation~\ref{eq:interfaceupdate} following Section~\ref{sec:levelset})
        \item[Step 4(f)] On $M_{\ell > 0}$ re-define cell types and unit normal and tangent vectors from updated level set.
        \item[Step 4(g)] Update to $t^{n+1}$ complete, $\forall M$ copy $\U_1\rightarrow\U_0$.
    \end{enumerate}
\end{enumerate}

\section{Results}
\label{sec:results}

In this section we present application of the method and verification tests for pure hydrodynamics, MHD, resistive MHD, two phase problems with different material properties on each side of the immersed boundary, and problems with a moving embedded boundary.
All results shown here were produced with the HLLD Riemann solver. Unless otherwise mentioned in the individual problem descriptions a $\gamma=5/3$ material is used.

\subsection{Standing acoustic wave in a cylinder}
\label{sec:wave_in_cylinder}
\begin{figure}[htb]
	\centering
	\subfloat[plot][Radial velocity]{%
		\includegraphics[width=.45\textwidth]{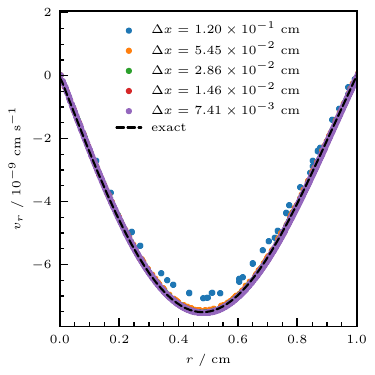}
	}
	\subfloat[plot][L1 error in density]{%
		\includegraphics[width=.45\textwidth]{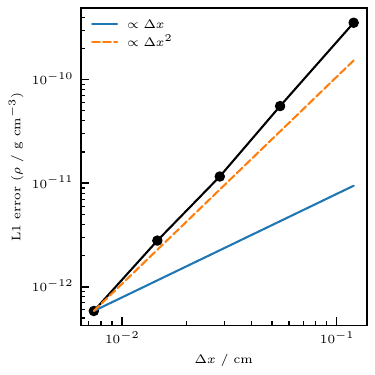}
	}
	\caption{Standing acoustic wave in a cylinder imposed using immersed
	boundary technique.
	}
	\label{fig:acoustic_cylinder}
\end{figure}

A standing acoustic wave is initialized inside of a cylindrical embedded boundary with radius $a=1$~cm. The background density and pressure are $\rho_0 = 1$~g~cm$^{-3}$ and $P_0 = 1$~Ba, on top of which a perturbation is imposed, giving an analytic solution as
\begin{align*}
    \rho_1 &= f\rho_0J_0(kr)\sin(\omega t) \ {\rm g\ cm^{-3}}\\
    P_1 &= \gamma P_0\dfrac{\rho_1}{\rho_0}, \\
    \rho &= \rho_0 + \rho_1 \\
    P &= P_0 + P_1 \\
    v_r &= -\dfrac{f\omega}{k}J_1(kr)\cos(\omega t)\ {\rm cm\ s^{-1}},
\end{align*}
where $\gamma=5/3$, $k=J_{11}/a$ is the wave number, $\omega=c_sk$ the frequency, $c_s = ({\gamma P/\rho})^{1/2}$ is the sound speed, $J_i$ is the $i^\mathrm{th}$-order Bessel function and $J_{11}=3.8317$ is the first zero of the first order Bessel function.
The perturbation magnitude is $f=10^{-8}$.
The embedded boundary is represented with a static polyline.
This gives an initial state with which to initialize the problem when $t=0$. The results are shown in Figure~\ref{fig:acoustic_cylinder} after one wave period and the solution converges at second order.

\clearpage
\subsection{Standing magnetosonic wave in a cylinder - perturbed $\theta$-pinch}
\label{sec:perturbed_theta_pinch}
\begin{figure}[htb]
	\centering
	\subfloat[plot][Axial field]{%
		\includegraphics[width=.45\textwidth]{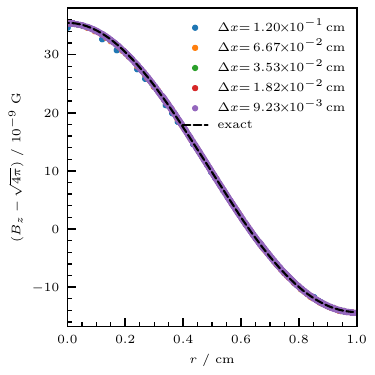}
	}
	\subfloat[plot][Error in density]{%
		\includegraphics[width=.45\textwidth]{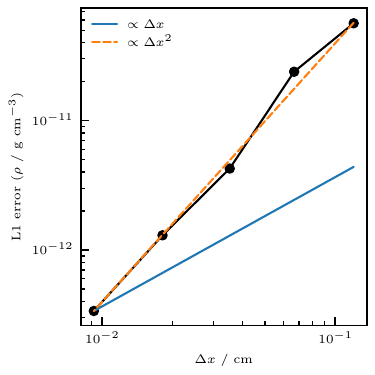}
	}

	\caption{Standing magnetosonic wave test in a cylinder using immersed
	boundaries. The simulation result converges to the exact analytic
	solution at second order. The setup is a perturbed $\theta$ pinch.
	}
	\label{fig:magnetosonic_standing_thetapinch}
\end{figure}

In this test, we extend the cylindrical acoustic wave problem by adding a magnetic field $B_{z,0}=\sqrt{4 \pi} $~G ($1$ in code units) in the $z$-direction, the analytic solution for which is then $B_z(t) = B_{z,0}(1+\rho_1(t)/\rho_0)$. The fast magnetosonic wave speed $c_p = ({c_A^2 + c_s^2})^{1/2}$ is used in place of the sound speed to calculate the frequency $\omega = c_pk$, where $c_A = ({B_z^2/\rho})^{1/2}$ is the Alfvén speed. The perturbation magnitude is again $f=10^{-8}$. 
The embedded boundary is represented with a polyline.
The results are shown in Figure~\ref{fig:magnetosonic_standing_thetapinch}  
after 7 oscillation periods
and the solution converges to the analytic one at second order.

\clearpage
\subsection{Standing magnetosonic wave in a cylinder - perturbed parallel pinch}
\label{sec:perturbed_parallel_pinch}
\begin{figure}[h!]
	\centering
	\subfloat[plot][Azimuthal Field]{%
		\includegraphics[width=.45\textwidth]{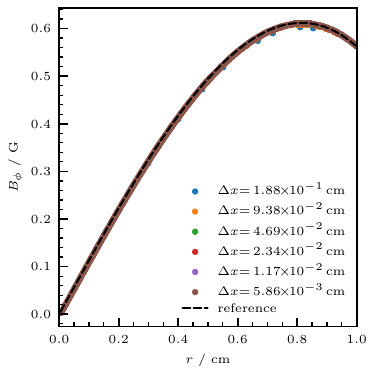}
	}
	\subfloat[plot][$B_\phi$ convergence]{%
		\includegraphics[width=.45\textwidth]{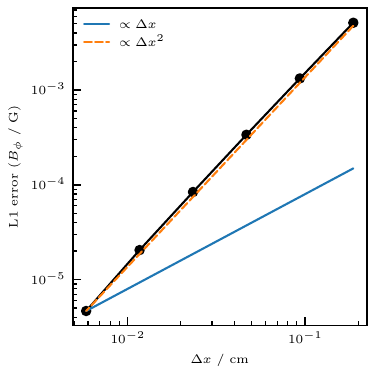}
	}
	\caption{Standing magnetosonic wave test in a cylinder using immersed
	boundaries. The setup is a perturbed parallel pinch. The gold solution
	that was used to compute the error was a high resolution simulation.
	}
	\label{fig:magnetosonic_standing_parallel}
\end{figure}

This test is again a standing magnetosonic wave in a cylinder but this time the configuration is a perturbed parallel pinch as opposed to just a $\theta$-pinch.
The parallel pinch magnetic field follows the form given in \citet[][eq.~5.33]{2014idmh.book.....F}.
We use the $z$-component of the vector potential to initialize the $x$ and $y$-direction magnetic fields, and the $z$ magnetic field directly.
In a cylinder of radius $a$ these are as a function of radial coordinate $r$:
\begin{align}
A_z(r) &= B_\phi(a) (a/4)\left(2-(r/a)^2\right)
\label{eq:pinchaz}\\
B_z(r) &= \frac{\sqrt{4 \pi}}{\alpha}\left[1+(2\alpha^2/3)(1-(r/a)^2)^2(5-2(r/a)^2)\right]^{1/2}\ {\rm G}
\label{eq:pinchbz}
\end{align}
where $\alpha=B_\phi(a)/B_z(a)$ and $B_\phi(a)$, $B_z(a)$ are given. For this test the density $\rho_0=1$, thermal pressure $P_0=6$~Ba, $a=1$~cm, $B_\phi(a)=\sqrt{\pi}/a$~G, and $B_z(a)=\sqrt{4 \pi}$~G.
The embedded boundary is represented with a polyline, and the overall mesh is $3$~cm square.
Results at $t=0.1$~s are shown in Figure~\ref{fig:magnetosonic_standing_parallel} demonstrate the self convergence of the solution from meshes $16^2$ to~$512^2$ with a $2048^2$ reference solution.

\clearpage
\subsection{Field line drag around perfectly conducting object}
\begin{figure}[htb]
 \centering
 \includegraphics[width=0.49\textwidth]{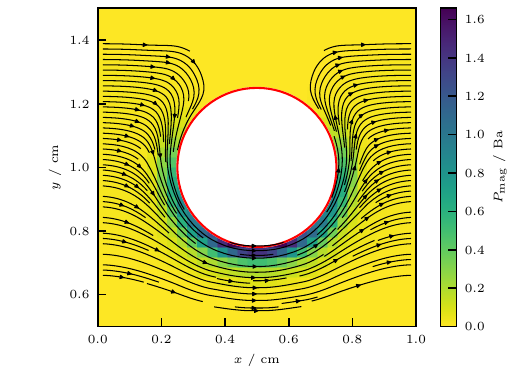}
 \includegraphics[width=0.49\textwidth]{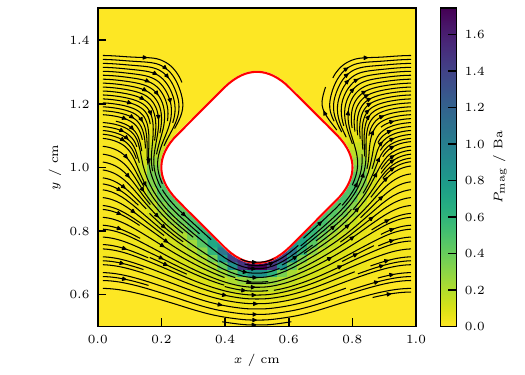}\\
  \includegraphics[width=0.49\textwidth]{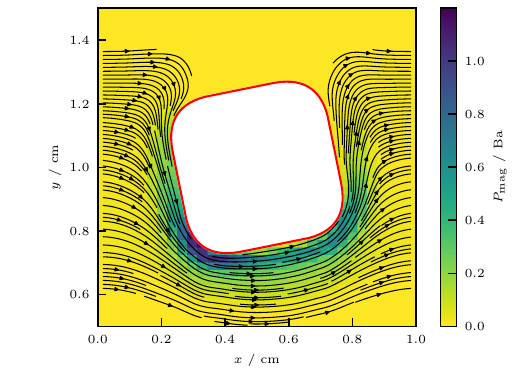}
 \caption{Ideal MHD simulation of the dragging of a magnetic field tangential to an
 immersed boundary with three different immersed boundary shapes. The domain shown is $32\times 32$ cells.
 The velocity and magnetic fields are initialy mutually perpendicular, with velocity in the $+\hat{y}$ direction and magnetic field in the $+\hat{x}$ direction.}
 \label{fig:fielddrag}
\end{figure}

Although in this qualitative problem there is no analytic solution available, it is designed to test the ability of the numerical method to deal with complex flow structures generated when a flow interacts with a perfectly conducting body. A fluid is initialized with uniform density ($\rho=1$~g~cm$^{-3}$) and pressure ($P_{\rm gas}=50$~Ba) and a rectangular horizontal band of magnetic field in the $x$-direction with  
a smooth profile as
\begin{align}
B_x(y) = \frac{\sqrt{4\pi}}{10}\cos\left(\frac{\pi}{8}\right)\left(1+\exp(-75(y-0.2))\right)^{-1}\left[1- \left(1+\exp(-75(y-0.6))\right)^{-1} \right]\, {\rm G}.
\end{align} 
The entire fluid has an initially uniform velocity in the $y$ direction ($v_y=0.75$~cm~s$^{-1}$), the domain has extents in $x,y$ of $[0,1],[0,2]$~cm with $32\times 64$ cells, and all domain boundaries are periodic.
As the setup is evolved, the flow should wrap around the body, and because we are assuming ideal MHD, the field should also wrap around the body. 
Ideally, no field lines should penetrate the perfect conductor boundary as they drag around the object, and the divergence-preserving nature of the constrained transport scheme preserves field lines in the active mesh.
However, in the cut cells, field lines can be advected so that they cross the boundary through the evolved faces of cut cells. Nevertheless, the divergence preserving quality of the scheme ensures that every flux line which enters the boundary still emerges from it in another location, so that the boundary behaves like one with a finite ``numerical'' resistivity due to the truncation error of the boundary scheme and no divergence appears in the magnetic field in the active domain.
We set up the problem with both a spherical object and a rectilinear object with rounded edges rotated with arbitrary angle to the mesh, both represented with a polyline. 
Both centered at $(0.5, 1.0)$~cm, the cylinder has radius $0.25$~cm, and the rounded square has side length $0.5$~cm with corners rounded to a radius of $0.15$~cm. 
We show results for variations where the rounded square is rotated $\pi/2$ and $\pi/16$.
In all cases, the same qualitative behavior is observed for a range of numerical resolutions. Results at time $t=0.8$~s are shown in Figure~\ref{fig:fielddrag}.

\clearpage
\subsection{Tangential field soak into conducting slab}
\begin{figure}[htb]
	\centering
	\includegraphics[width=0.45\textwidth]{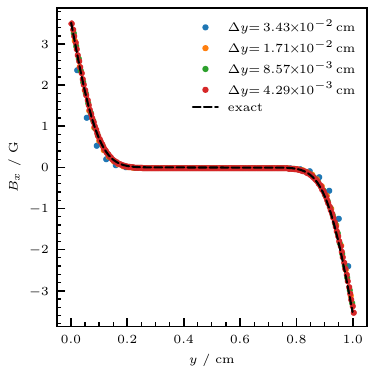}
	\includegraphics[width=0.45\textwidth]{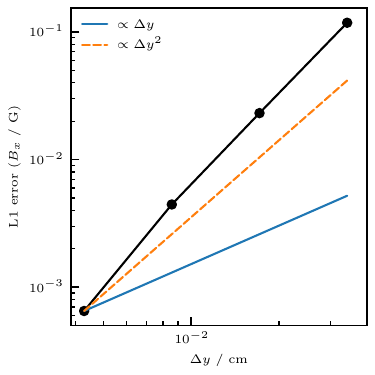}
	\caption{Tangential field soak into a conducting slab from immersed
	boundaries at $y=0$ and $y=1$. Initially there
	is no flux density in the domain $0 < y < 1$. Left panel: $B_x$ is shown at the end
	of the calculation at time $t=3\times10^{-6}$~s. Right panel:
	convergence to the semi-analytic solution is shown to be second order.}
	\label{fig:bxslabsoak}
\end{figure}

To verify the resistive operator, we produce a test with locally essentially one-dimensional behavior which can be compared to an analytical solution. A conducting square 
(side length $1$~cm) 
with resistivity $10^3$~cm$^2$~s$^{-1}$~in a 
$1.2$~cm 
domain containing initially no magnetic flux density is subjected to an external field along its boundary with $\B\cdot\tang = \sqrt{4 \pi}$~G, $\B\cdot\norm=B_z=0$, imposed using a Dirichlet condition for the embedded boundary,
which is represented with a polyline.
An accurate semi-analytical solution for the tangential field in the interface-normal direction
far from the corners (e.g.\ the side midpoints)
can be obtained for some periodic domain $0<y<L$ at short times is a sum of error functions:
\begin{equation}
    \frac{B(y,t)}{\sqrt{4\pi}}= 1 - (-1)^N - \sum_{n=-(N-1)}^N (-1)^n\erf\left(\frac{y-nL}{\sqrt{4\eta t}}\right)\ {\rm G}.
\end{equation}
This approximation works well as long as the diffusion length $(\eta t)^{1/2}< L/2$ and only requires $N=6$ to reach machine precision at this time.  At longer times the accuracy degrades and it is better to use the Fourier series
\begin{equation}
    \frac{B(y,t)}{\sqrt{4\pi}} =1-\frac{4}{\pi}\sum_{n=1,3,5,\ldots}^\infty \sin\left(\frac{n\pi y}{L}\right)e^{-(n\pi/L)^2\eta t}\ {\rm G}
\end{equation}
which converges rapidly (two terms yields machine precision at these longer times).
We use the solution from the former series summation to compute the L1 error in the numerical result at $t=3\times10^{-6}$~s, which is shown in Figure~\ref{fig:bxslabsoak} to converge at second order.

\subsection{Relaxation of axial field in cylindrical conductor}
\label{sec:cylbzsoak}
\begin{figure}[htb]
	\centering
	\includegraphics[width=0.55\textwidth]{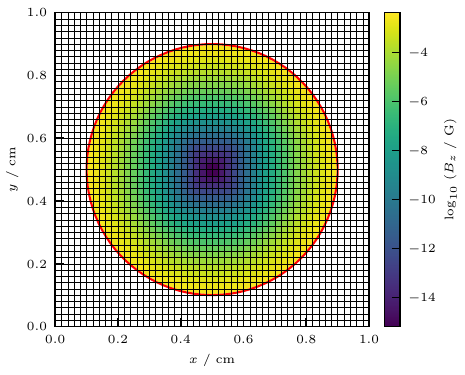}
	\includegraphics[width=0.44\textwidth]{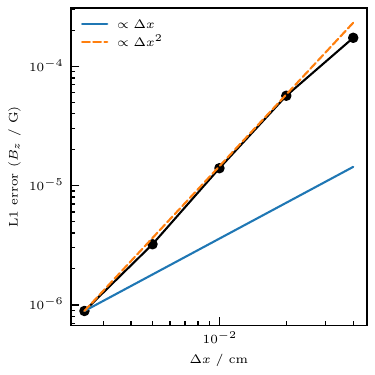}
	\caption{Resistive transport of axial magnetic flux inwards from the
	edge of a cylindrical conductor.
	Left: $50\times50$ simulation.
    Right: Convergence of the solution in the axial field
	soak problem. The error is computed by comparing to an analytic
	solution.  Solid blue ($\mathcal{O}(\Delta x)$) and dashed orange
	($\mathcal{O}(\Delta x^2)$) reference lines are given for comparison.}
	\label{fig:cylinderzsoak}
\end{figure}

This problem simulates the resistive transport of axial magnetic flux from
the outer boundary of a cylindrical conductor towards its center. Initially,
there is no field inside the conductor, and on the boundary we have $B_z =
\sqrt{4 \pi} 10^{-3}$~G. Magnetic diffusivity is assumed to be spatially uniform at $\eta=10^{-3}\ \mathrm{cm^2 s^{-1}}$, the conductor has radius $a=0.4$
and the results are compared at a final time of 1~s. 
The semi-analytical solution is given in terms of a Fourier-Bessel series:
\begin{equation}
    B_z(r,t)=B_{z}(a)\left[1-\sum_{n=1}^\infty c_n J_0(k_n r) e^{-\eta k_n^2 t}\right]
\end{equation}
Here the $k_n$ are the positive roots of $J_0(k_na)=0$, where $a$ is the radius of the cylinder.
The coefficients $c_n=2/k_naJ_1(k_na)$ are chosen to satisfy the initial condition $B_z(r,0)=0$ for $r<a$.
The embedded boundary is represented with a polyline.
The convergence of the solution, shown in Figure~\ref{fig:cylinderzsoak},
is of second order.

\subsection{Relaxation of azimuthal field in cylindrical conductor}
\label{sec:cylbphisoak}
\begin{figure}[htb]
	\centering
    \includegraphics[width=0.54\textwidth]{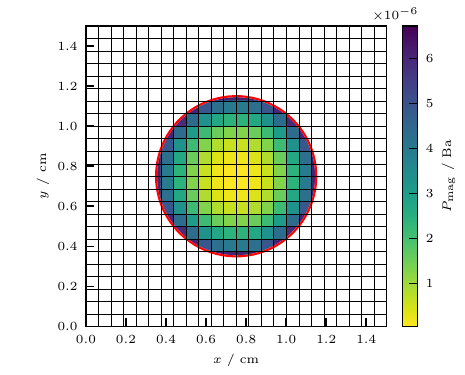}
    \includegraphics[width=0.44\textwidth]{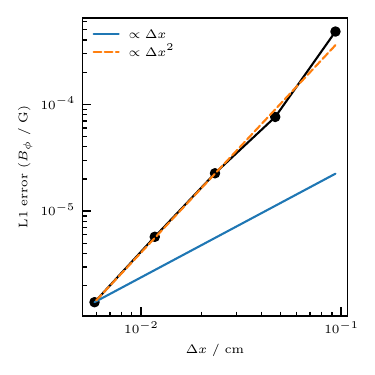}
	\caption{Resistive transport of toroidal magnetic flux inwards from the
	edge of a cylindrical conductor.
    Left: $24\times24$ simulation.
    Right: Convergence of the solution in the toroidal field
	soak problem. The error is computed by comparing to a analytic
	solution.  Solid blue ($\mathcal{O}(\Delta x)$) and dashed orange
	($\mathcal{O}(\Delta x^2)$) reference lines are given for comparison.
	}
	\label{fig:cylinderbphisoak}
\end{figure}

Similarly to the problem in \ref{sec:cylbzsoak}, we initialize a cylindrical
conductor with zero magnetic flux density, and impose a toroidal field with
$B_\phi=\sqrt{4\pi} 10^{-3}$~G via the immersed boundary condition. The result at $t=0.03$~s with a
resistivity of $\eta=1\ \mathrm{cm^2 s^{-1}}$ and a conductor with radius $a=0.4$~cm is shown in
Figure~\ref{fig:cylinderbphisoak} for a series of resolutions. 
The semi-analytical solution has a Fourier-Bessel series in $J_1$ for the deviation of the solution from the asymptotic $t\rightarrow\infty$ solution $B_{\phi}(a)r/a$:
\begin{equation}
    B_\phi(r,t)=B_{\phi}(a)\left[r/a-\sum_{n=1}^\infty c_n J_1(k_n r) e^{-\eta k_n^2 t}\right]
\end{equation}
Here $k_n$ are the non-trivial positive roots of $J_1(k_n a)=0$, where $a$ is the radius of the cylinder.  The coefficients $c_n=2/k_naJ_2(k_na)$ are chosen to satisfy the initial condition $B_\phi(r,0)=0$ for $r<a$.
The embedded boundary is represented with a polyline.
The L1 error norm
computed by comparing with the analytic solution is shown to converge at
second order in Figure~\ref{fig:cylinderbphisoak}.

\clearpage
\subsection{Resistive magnetic field soak from one material slab to another}
\label{sec:slabsoak}
\begin{figure}[htb]
	\centering
	\includegraphics[width=0.49\textwidth]{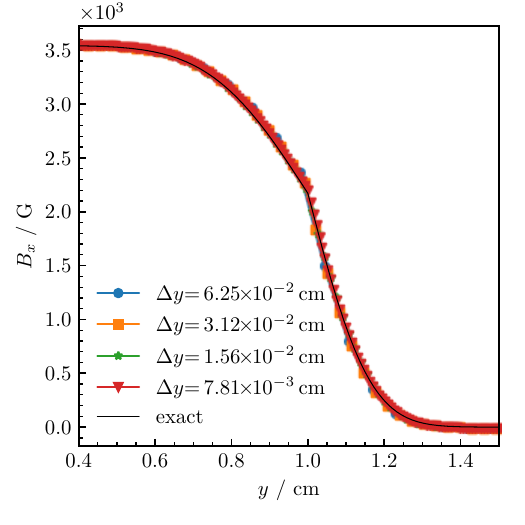}
	\includegraphics[width=0.49\textwidth]{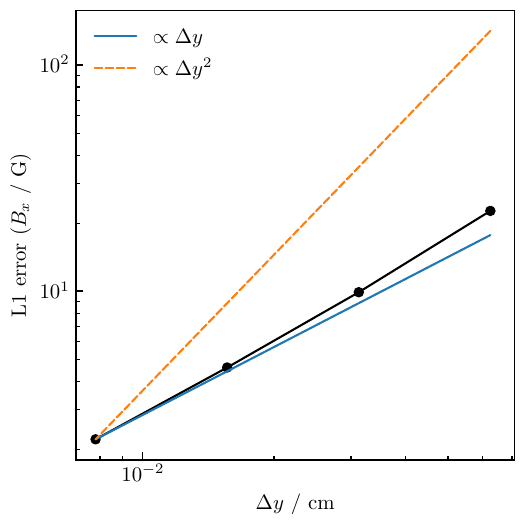}
	\caption{Convergence of the solution in the two-material slab field
	soak problem. The error is computed by comparing to an analytic
	solution (labeled `exact').  $\mathcal{O}(\Delta y)$ and
	$\mathcal{O}(\Delta y^2)$ reference lines are given for comparison.
	}
	\label{fig:twosided_slab_soak}
\end{figure}

A uniform density and pressure slab is initialized with two adjacent materials
with the same density $\rho=10^3$~g~cm$^{-3}$ in mechanical equilibrium with
pressure $P=6\times10^6$~Ba. One material has an initial uniform magnetic field $B_x=\sqrt{4 \pi}10^3$~G and magnetic diffusivity of
$1\times10^7\ \mathrm{cm^2 s^{-1}}$ and the other has a magnetic diffusivity of $4\times10^6\ \mathrm{cm^2 s^{-1}}$ and zero initial magnetic field.
The domain has extents $0.125$~cm by $1$~cm with zero gradient (Neumann) boundary conditions and the interface is described with a static level set. 
To exercise the cut-cell method, care is taken in the resolution study to set the material interface offset by a fraction of a cell from the cell face, by setting it at $y=1+sin(1)\Delta x$.
For ease of interpretation we shift the results to show the interface at $y=1$~cm below,
Results at $t=2\times10^{-9}\ \mathrm{s}$ are shown in Figure~\ref{fig:twosided_slab_soak}.
The convergence rate of the L1 error of the cell-averages is expected to be first order due to the discontinuity in the resistivity, and this is reflected in the numerical results.

\subsection{Resistive magnetic field soak across cylindrical material interface}
\begin{figure}[htb]
	\centering
	\subfloat[plot][$B_\phi$ result]{
        \includegraphics[width=0.49\textwidth]{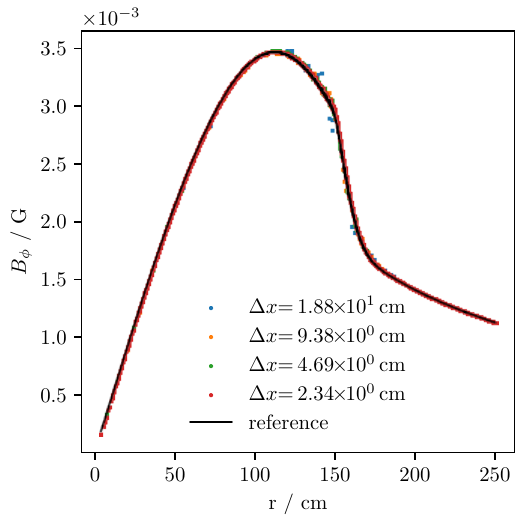}
    }
    \subfloat[plot][Self-convergence]{
	    \includegraphics[width=0.49\textwidth]{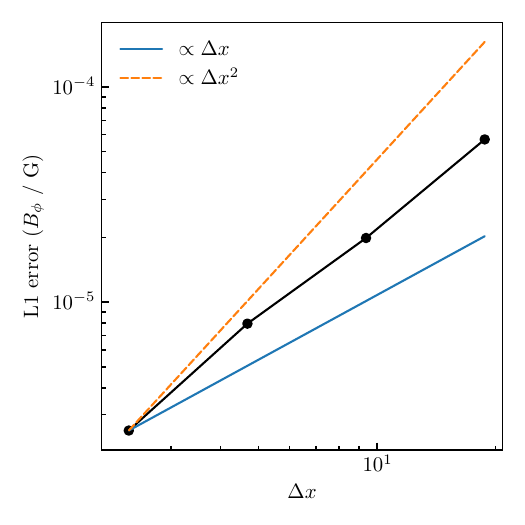}
    }\\
    \subfloat[plot][$B_\phi$ for $t=5\times10^{-2}\ \mathrm{s}$ at $128^2$]{
	    \includegraphics[width=0.49\textwidth]{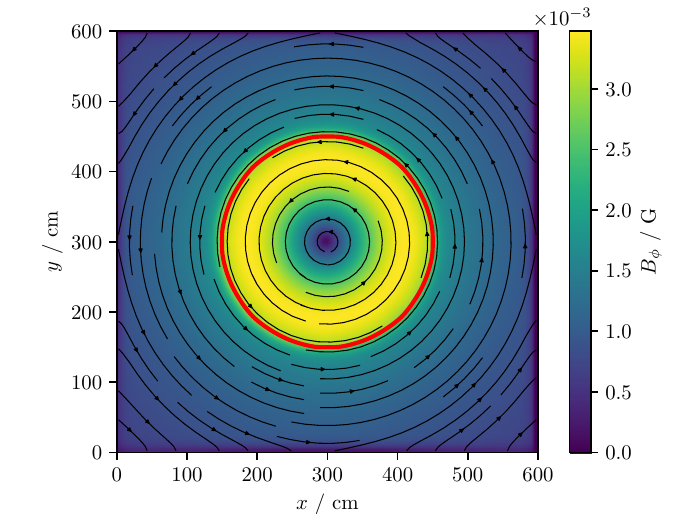}
    }
    \subfloat[plot][$\nabla\cdot\B$ at $128^2$]{
         \includegraphics[width=0.49\textwidth]{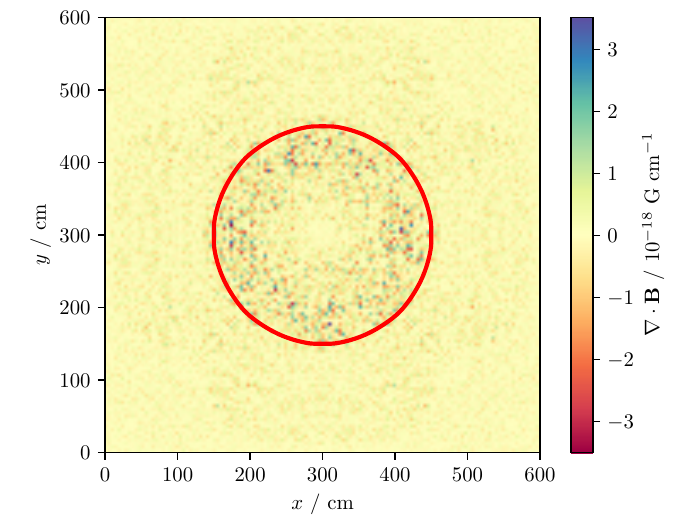} 
    }
	\caption{Result (top left) and self-convergence analysis (top right) for a two-material cylinder problem.
    Final $B_\phi$ (bottom left) and $\nabla\cdot\mathbf{B}$ (bottom right) on a $128^2$ mesh are also shown.
	}
	\label{fig:twosided_cylinder_soak}
\end{figure}

This problem demonstrates the diffusion of a magnetic field across a curved interface between two phases with differing properties.
To create a field with a sharp gradient at the interface, we employ the parallel pinch fields as in Section~\ref{sec:perturbed_parallel_pinch}, but rescale the vector potential (\ref{eq:pinchaz}) and $z$-field (\ref{eq:pinchbz}) inside the pinch radius $a$ by $1/2$, but use the vacuum field for a conductor carrying the current corresponding to the unscaled solution at radii greater than $a$.
Thus, by placing the material interface at the same location, a initial condition with a sharp discontinuity at this cylindrical material interface is produced.
We set this pinch and interface radius to  $a=150\ \mathrm{cm}$ as shown in Figure~\ref{fig:twosided_cylinder_soak}.
Inside the cylindrical interface, the magnetic diffusivity is set as $10^4\; {\rm cm^{2}\, s^{-1}}$ and $10^3\; {\rm cm^{2}\, s^{-1}}$ outside. 
On the edges of the square computational domain, all boundaries are periodic. 
The small magnitude of the resistivity in the outer region prevents significant reconnection of the magnetic field at these boundaries during the timescale analyzed.
The interface here is stationary, and represented with a level set.
The setup is evolved for $5\times10^{-2}\ \mathrm{s}$.
Like in the slab geometry test (Section~\ref{sec:slabsoak}), the discontinuity in resistivity is expected to result in first-order convergence. The self-convergence rate (calculated with a result of $512^2$ as reference) shown in Figure~\ref{fig:twosided_cylinder_soak}(panel b) is between first and second order, suggesting that the first-order nature of the scheme at the discontinuity is not yet dominating the total L1 error at the time analyzed.
Finally, this test demonstrates that the magnetic divergence errors are maintained to within machine precision in a two-sided interface problem with non-mesh-aligned interface as shown in Figure~\ref{fig:twosided_cylinder_soak}(panel d).

\clearpage
\subsection{Compression of a $\theta$--pinch}
\label{sec:theta_pinch_compression}
\begin{figure}[tb]
	\centering
	\subfloat[plot][Density]{%
		\includegraphics[width=.49\textwidth]{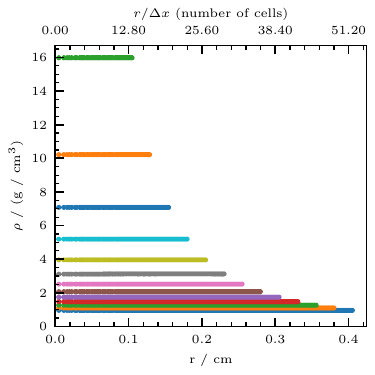}
	}
	\subfloat[plot][Axial field]{%
		\includegraphics[width=.49\textwidth]{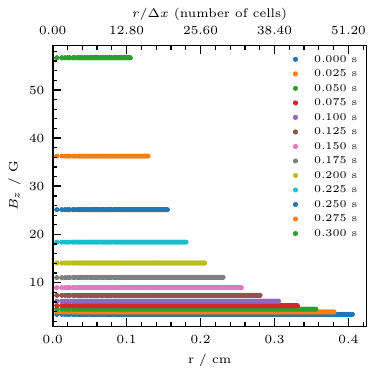}
	} \\
	\subfloat[plot][Central pressure and density]{%
		\includegraphics[width=.4\textwidth]{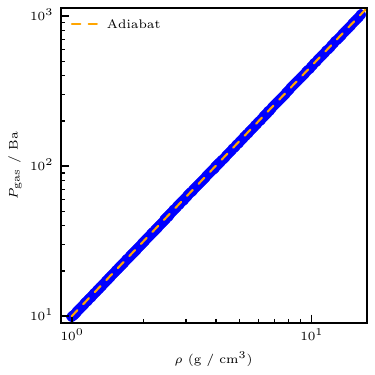}
	}
	\subfloat[plot][Conservation error]{%
		\includegraphics[width=.4\textwidth]{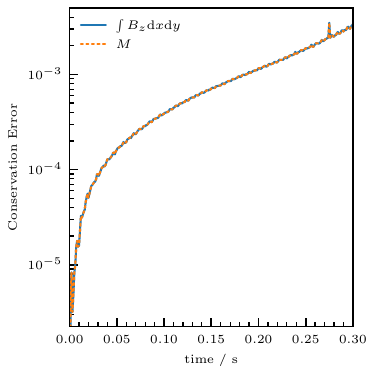}
	}

	\caption{Sub-Alfv\'enic compression of a theta pinch configuration
	using a moving boundary. Initially, there are 24 grid cells over
	the radius of the cylinder, decreasing throughout the compression
	simulation.
	}
	\label{fig:theta_pinch_compression}
\end{figure}

To demonstrate application of the method to a problem with a moving boundary, we present here a pair of simple setups capturing the adiabatic heating of a plasma by a compressing conducting wall.
The first is the compression of a $\theta$-pinch, in the form of a plasma with a uniform magnetic field in the $z$ direction.
Inside a square domain of side length unity, on a Cartesian mesh of resolution $128^2$, a cylindrical embedded boundary with radius $a=0.4$~cm was initialized.
This gives an initial resolution of 51.2 cells in the initial radius of the cylindrical domain, decreasing over time, and MHD equations are solved only inside of this boundary.
The magnetic field is set as uniform $B_z = \sqrt{4 \pi}$~G
and the velocity is set as a flow converging to the center of the cylindrical domain with radial velocity
$v_{\rm r} = -r/a$.
A uniform density of $\rho=1$~g~cm$^{-3}$ and thermal pressure $P_{\rm gas}=10$~Ba was used.
The cylindrical boundary is a reflecting perfect conductor, and moves inwards following the same velocity profile as the initial plasma.
The embedded boundary is represented with a moving polyline with specified motion following $v_{\rm r} = -r/a$.
The fluid magnetic pressure is subdominant and the motion of the wall is thus strongly strongly sub-Alfvenic, leading to a smooth, uniform density converging flow at quasi-uniform density.
As there are no fluxes across the compressing boundary the density and magnetic field rise during the compression adiabatically.
The results of this $\theta$-pinch compression are shown in 
Figure~\ref{fig:theta_pinch_compression}.
Note in the plots a point is shown to represent a value at the geometric center of the entire mesh cell, including for cut cells where only the part of the cell is active.
The central pressure and density in the cylindrical domain evolves adiabatically, increasing in density, pressure and temperature (Figure~\ref{fig:theta_pinch_compression}, panel c).
The magnetic field $B_z$ is also increased by the compression. However, there is some conservation error, both in the total magnetic flux and the total mass inside the compressing boundary.
These have nearly the same relative error (Figure~\ref{fig:theta_pinch_compression}, panel d), as both are essentially solved in very similar ways in this two dimensional implementation. 
The in-plane components of the magnetic field are evolved differently, so a similar test with a parallel pinch in the next subsection will display different behavior.

\clearpage

\subsection{Compression of a parallel--pinch}
\begin{figure}[h]
	\centering
	\subfloat[plot][Density]{%
		\includegraphics[width=.33\textwidth]{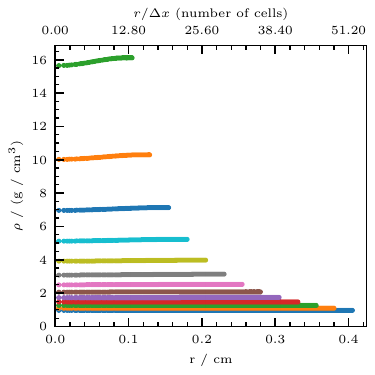}
	}
	\subfloat[plot][Azimuthal field]{%
		\includegraphics[width=.33\textwidth]{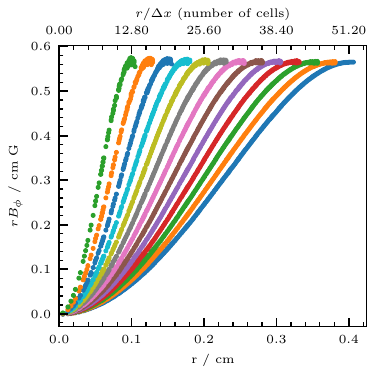}
	}
	\subfloat[plot][Axial field]{%
		\includegraphics[width=.33\textwidth]{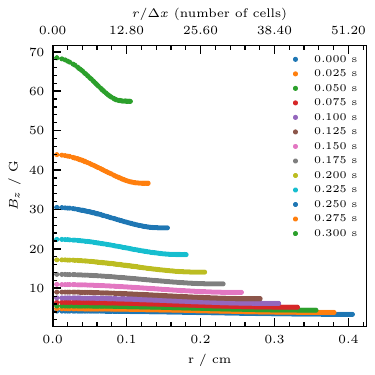}
	} \\
	\subfloat[plot][Central pressure and density]{%
		\includegraphics[width=.33\textwidth]{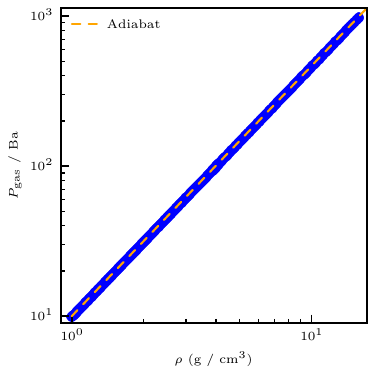}
	}
	\subfloat[plot][Conservation errors]{%
		\includegraphics[width=.33\textwidth]{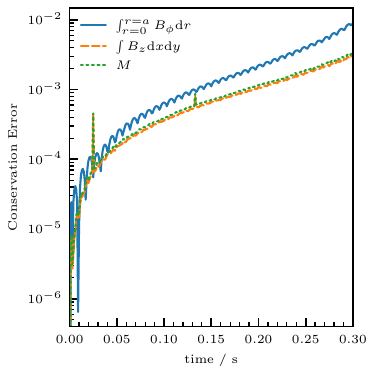}
	}
	\caption{Sub-Alfv\'enic compression of a parallel pinch configuration
	using a moving boundary. Initially, there are about 48 grid cells over
	the radius of the cylinder, decreasing throughout the compression
	simulation.
	}
	\label{fig:parallel_pinch_compression}
\end{figure}

\FloatBarrier
A second version of the cylindrical compression test uses a parallel pinch magnetic field configuration, the equilibirium solution previously described in Section~\ref{sec:perturbed_parallel_pinch}
with $B_\theta(a)=\sqrt{\pi}/a$~G, $B_z(a)=\sqrt{4 \pi}$~G.
Otherwise all parameters and boundary conditions are the same as in the $\theta$-pinch test of Section~\ref{sec:theta_pinch_compression}, and again the embedded boundary is represented with a polyline.
Results of this compression test are shown in Figure~\ref{fig:parallel_pinch_compression}.
Again, the strongly sub-Alfvenic compression results in a smooth, nearly adiabatic compression of the initial condition.
However, the internal structure of the equilibrium changes as it is compressed, resulting in evolution of the radial profiles of density and magnetic field.
Again, the global non-conservation errors in total mass and $z$-direction magnetic flux are very similar early (e.g.\ at high resolutions) but diverge later (e.g.\ at low resolutions). They are now also non-monotonic in magnitude.
As the in-plane ($B_\theta$) magnetic field is non-zero in this configuration, a conservation error of this component of magnetic flux due to the boundary condition is again present.

\clearpage
\subsection{Hydrodynamic shock compression of a parallel pinch with material interface}
\label{sec:shock-converge}
\begin{figure}[tb]
    \centering
    \includegraphics[width=\textwidth]{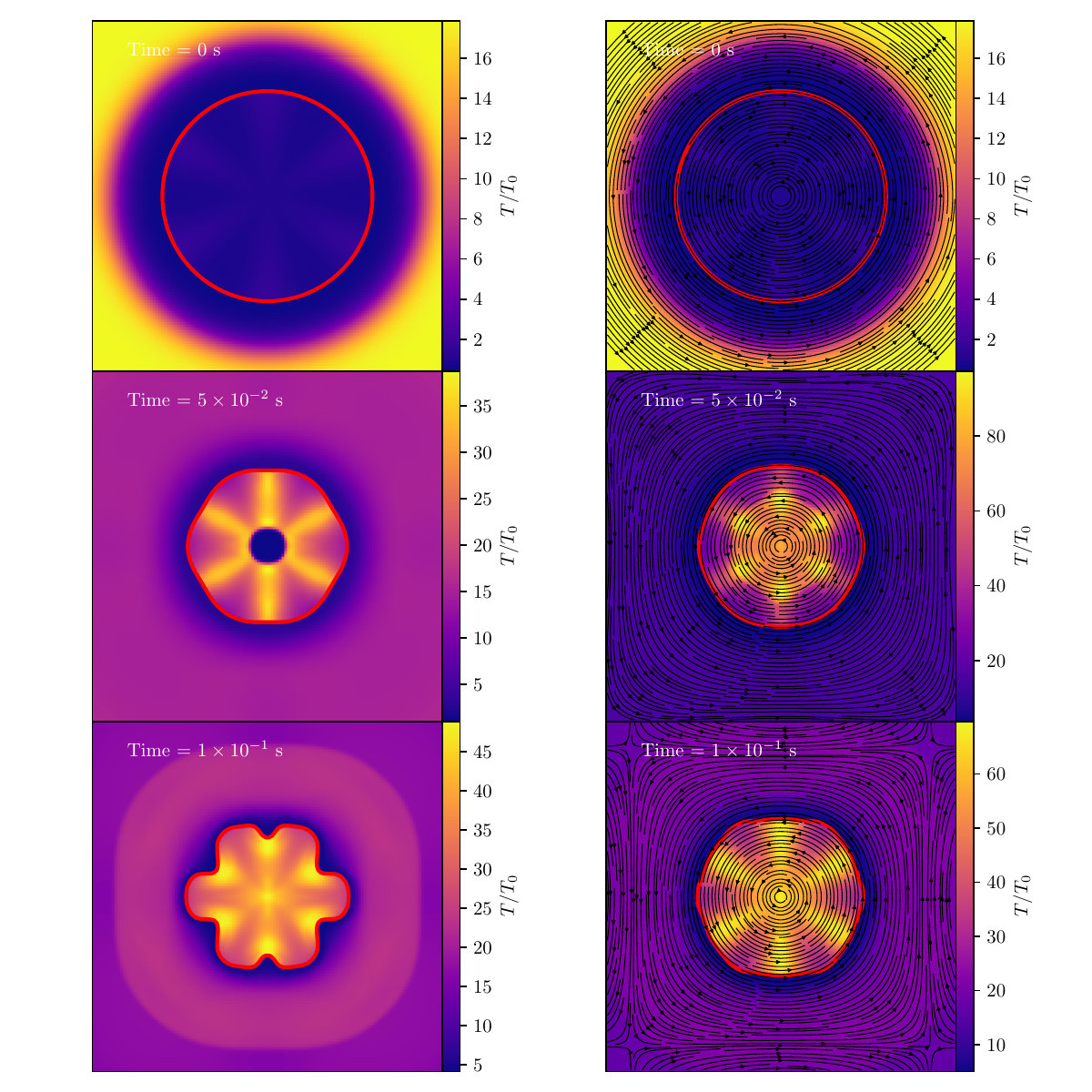}
    \caption{Converging shock compressing a parallel pinch. {\sl Left:} Unmagnetized {\sl Right:} Magnetized. Domain is a $1~\mathrm{cm}\times1~\mathrm{cm}$ square.}
    \label{fig:shock-converge-pinch}
\end{figure}
This problem consists of an infinite outer cylinder containing a material with $\gamma_1=5/3$ and initial density $\rho_1=1.4$~g~cm$^{-3}$ surrounding an inner, second cylinder with radius $r_2=0.3$~cm containing a second material with $\gamma_2=4$ and $\rho_2=0.5$~g~cm$^{-3}$.
The density in the inner cylinder contains a sinusoidal perturbation in $\phi$ with amplitude $0.1$, and the perturbation decays towards $r=0$~cm. A complete expression for the density of the inner material is
\begin{align*}
    \rho_2 = \frac{1}{2} + 0.1\sin\left(6\phi+\frac{3\pi}{2}\right)
    \left[1-\cos\left(\frac{\pi r}{r_2}\right)\right] ~{\rm{g~cm^{-3}}}.
\end{align*}
Both materials are initialized in mechanical equilibrium with $P_\mathrm{gas}=1$~Ba at the material interface, and a converging shock is driven inwards by initializing a region in material 1 with $P=50$~Ba for $r>0.45$~cm. A zero-gradient Neumann boundary condition is used at the simulation domain boundary. We perform two versions of the test -- one with no magnetic fields and one with magnetic fields. In the case with magnetic fields, the materials have magnetic diffusivities $\eta_1=1$~cm$^2$~s$^{-1}$ and $\eta_2=0.1$~cm$^2$~s$^{-1}$. The magnetic field is initialized using the parallel pinch vector potential and axial field given in equation~(\ref{eq:pinchaz}--\ref{eq:pinchbz}) with $a=r_2$, $B_\phi(a)=5\sqrt{\pi}/a$~G and $B_z(a)=0.1\sqrt{4\pi}$~G.
The interface is tracked with a level set.
Three snapshots in time are shown in Figure~\ref{fig:shock-converge-pinch} for both simulations. In both cases, the hydrodynamic shock compresses the inner material, which heats up. The shock reflects off the axis of the cylinder and the reverse shock strikes the interface a second time. In the case without magnetic fields, the Richtmyer-Meshkov instability grows from the density perturbation seeds. In the case with magnetic fields, the instability is suppressed.

\clearpage
\subsection{Shocked bubble with material interface}
\begin{figure}
    \centering
    \includegraphics[width=\textwidth]{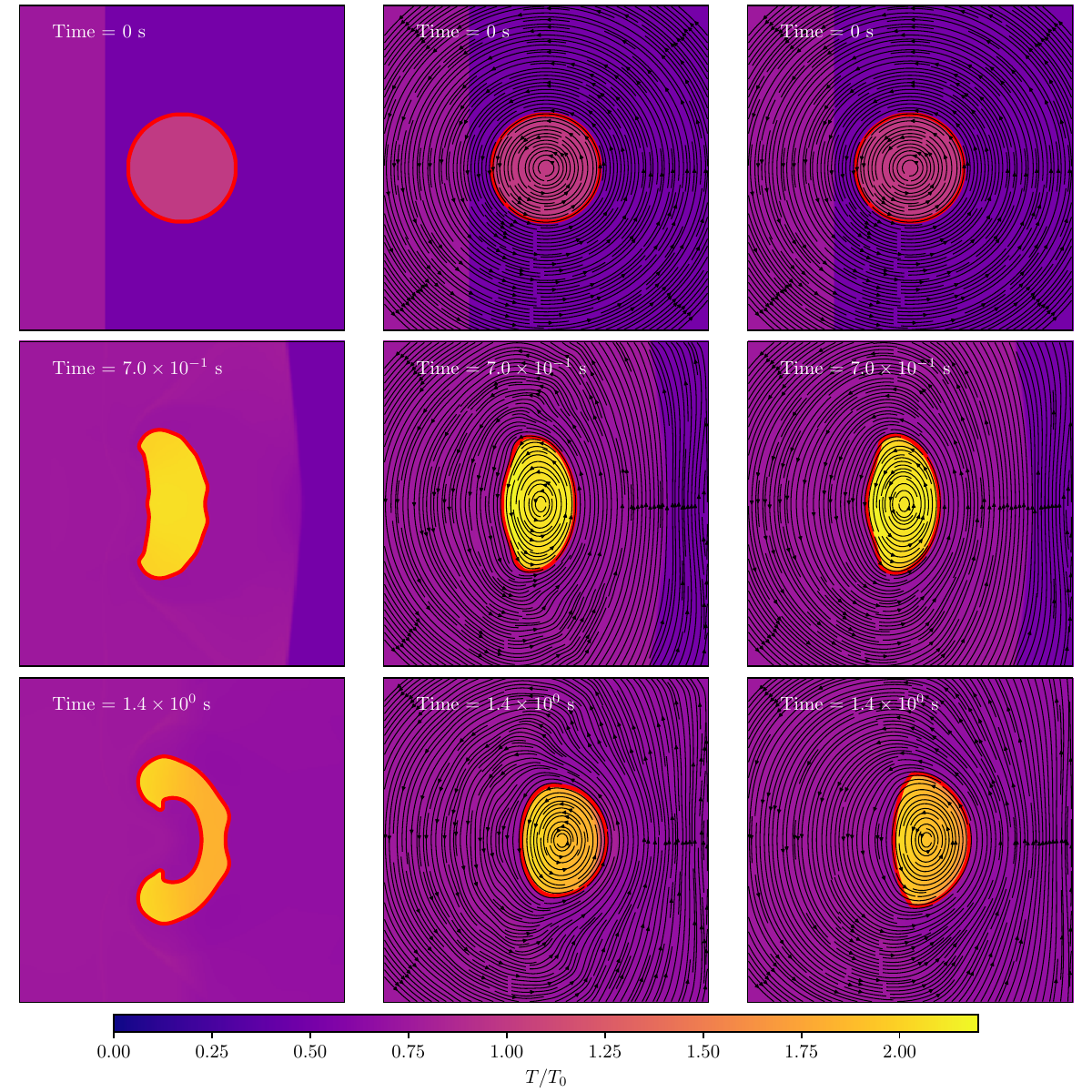}
    \caption{Shocked bubble problem with material interface. Left: unmagnetized. Middle: magnetized, ideal MHD Right: magnetized, finite resistivity. Domain is $1.5~\mathrm{cm}\times1.5~\mathrm{cm}$ with a $200 \times 200$. cell mesh $T_0$ is the initial temperature in the bubble.}
    \label{fig:shock_bubble}
\end{figure}

This setup is based on the classic shocked helium bubble problem \citep[e.g.][]{quirk1996a}, 
but here with a variation unconstrained by physical experiment and which demonstrates the abilities of the scheme. A cylindrical bubble with density $\rho_2=0.5$~g~cm$^{-3}$, radius $r_2=0.25$~cm, adiabatic index $\gamma_2=5$, and centered at $(x,y)=(0.75,0.75)$~cm is initialized in pressure equilibrium with a background material with $\rho_1=2$~g~cm$^{-3}$, $\gamma_1=5/3$, and $P=1$~Ba. A planar shock is initialized with the hydrodynamic jump conditions for a mach 1.5 shock at $x=0.4$~cm, and the flow is initially set in a frame where the post-shock flow velocity is zero, so that the bubble generally stays within the domain.
For consistency, this same initialization is used in variations of this test with and without a magnetic field.
The shock is driven to the right through the background material from $x = 0.1$~cm, impinging upon the bubble and accelerating it to the right from rest. In the magnetized version of this problem, the field is initialized as in the converging shock problem described in Section~\ref{sec:shock-converge}, with $a=r_2=0.25$~cm, 
$B_\phi(a)=\sqrt{\pi}/ a$~G
and $B_z(a)=0.1\sqrt{4 \pi}$~G.
In the nonideal MHD variation with finite resistivity, the materials have magnetic diffusivities $\eta_1=1$~cm$^2$~s$^{-1}$ and $\eta_2=0.1$~cm$^2$~s$^{-1}$.
The interface is tracked with a level set.
In the unmagnetized version, the hydrodynamic shock deforms the bubble in the same qualitative manner as has been found in both experiment and simulation in the aforementioned literature. In the magnetized case, the pinch equilibrium pushes back and prevents the bubble from deforming as readily, as we show in Figure~\ref{fig:shock_bubble}. 
The variations with ideal and resistive MHD demonstrate the ability of the scheme to capture the complex interplay between the multiple physical effects.
\clearpage
\subsection{Magnetic compression of a z-pinch with 
material interface}
\label{sec:maglif}

\begin{figure}[htb]
    \centering
    \includegraphics[width=\linewidth]{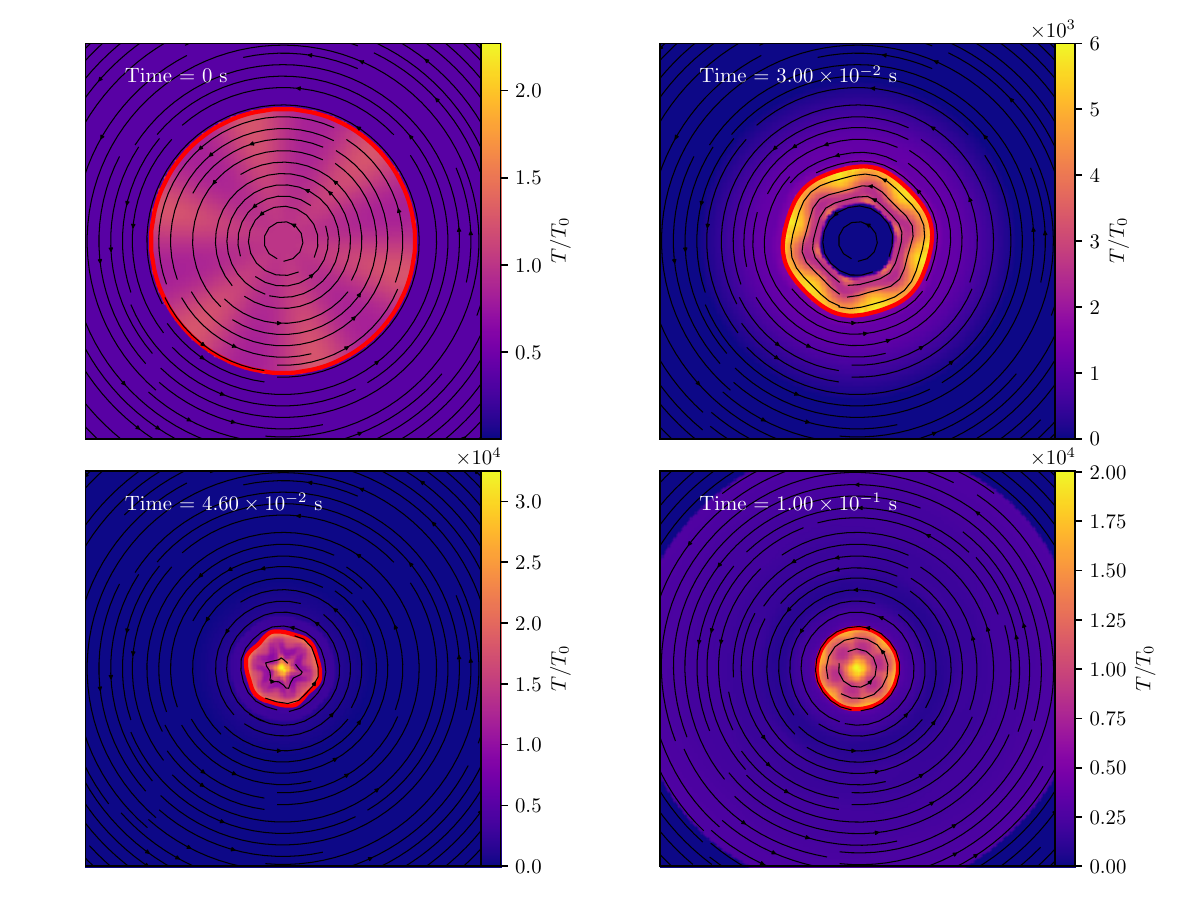}
    \caption{Magnetic compression of a parallel pinch. The total simulation domain is $32~\mathrm{cm}\times32~\mathrm{cm}$ with a $512\times512$ mesh, but here only the central $12$~cm section is shown to highlight the compression.}
    \label{fig:maglif}
\end{figure}

The compression of a cylindrical liner by a magnetic field is a useful application of the present method. In this test, we demonstrate our algorithm applied to such a problem. 
The material inside the liner has $\gamma_1=2$, $\eta_1=0.1$~cm$^2$~s$^{-1}$ and is initialized with density perturbations as
\begin{align*}
    \rho_1 = \frac{1}{2} + 0.05\sin(6\phi)
    \left[1-\cos\left(\frac{\pi r}{r_1}\right)\right]~{\rm g\ cm^{-3}}.
\end{align*}
where the liner inner radius is $r_1=4$~cm and this is the initial location of the material interface. 
The liner material (at $r>r_1$) has $\gamma_2=1.3$, $\eta_2=0.05$~cm$^2$~s$^{-1}$ and $\rho_2=1.4$~g~cm$^{-3}$ except at $r>3r_1$ where we set the density $100\rho_2$ to provide a buffer region between the computationally expedient periodic boundary condition applied to the square edges of the domain and the inner section with the circular pinch compression.\footnote{Extending the implementation to support both a one-sided static boundary here and the moving material interface in the same computation would be another valid approach to this problem setup.}
Both materials are initialized with $P_\mathrm{gas}=1$~Ba. $B_z=0$ everywhere, and the azimuthal field is initially
\begin{align*}
    B_\phi = -100\sqrt{4\pi}\left(1 + \tanh\left[\frac{r-a}{2}\right]\right)\ {\rm G},
\end{align*}
resulting in an outwardly pointing magnetic pressure gradient and is initialized using the vector potential
\begin{align*}
    A_z = -200\sqrt{4 \pi}\left(
    \frac{r}{2} + \log\left[\cosh\left(\frac{a-r}{2}\right)\right]
    \right)\ {\rm G\ cm}.
\end{align*}
We use $a=5$~cm.
The interface is tracked with a level set, and all domain boundary conditions are zero-gradient.
A time series of the simulated resulting magnetically-driven compression is shown in Figure~\ref{fig:maglif}.

\section{Conclusions}
\label{sec:conclusions}
We have presented a method for implementing a moving embedded boundary into a finite-volume constrained-transport solution of the resistive MHD equations.
Here we presented results for one-sided moving and fixed boundaries, and two-sided fixed and moving boundaries.
We have also only presented methods for two spatial dimensions.
Further extension to three dimensions should be straightforward and would expand the regime of application these techniques significantly.
It is helpful that the building blocks of the scheme have previously been developed in three dimensions: The \cite{forrer1998a} cut-cell method has previously been implemented in three dimensions with non-constrained-transport methods for the induction equation \citep{2017PhPl...24b2501L,2018PhPl...25l2513L}, and the ghost fluid and level set methods are extensively developed in three dimensions \citep{osher2004a}.
Only the extensions for constrained transport MHD described in Section~\ref{sec:mhd_extension}, require new adaption, and this is primarily a matter of removing the special case of the $B_z$ component and treating it equivalently to the $x$ and $y$ components of the magnetic field.
For treatment of systems with dominant axisymmetry, logically Cartesian but geometrically non-Cartesian meshes (such as cylindrical polar) are of great interest.

The underlying method for handling the continuity, momentum, and energy equations at the boundary used here is the \citet{forrer1998a} cut-cell method, which is not fully conservative. 
Other cut cell methods such as the flux-redistribution scheme \citep{chenandcolella87} can avoid the non-conservation at the cost of additional complexity.

Our method for solving the induction equations is to apply the electric fields from each side of the interface to update a single global \B field, represented by face averages in the underlying mesh.
This implies some numerical diffusion of magnetic field across the interface, which may be problematic in some situations where the consequences of a large jump in resistivity between the phases is important to the physical situation being modeled.
In our experimentation we also found it was possible to apply a constrained-transport update at the boundary cut cells in a way that maintains the divergence-free property of the solution in the interior of each phase, but does not conserve magnetic flux at the interface between two phases.
This can minimize the numerical diffusion, but at the cost of a mismatch between the two sides of the interface.
A cut-cell method which can accomplish both these goals at the same time would be a significant step forward from this work. 
A moving unstructured mesh with faces conformal to the interface would also provide a route around this difficulty, although at the cost of several other complications and sources of numerical dissipation.

The requirement that the virtual cell resulting from reflection of a ghost cell in the boundary lie within the active domain places a limit on the curvature of the boundary. 
However, as this scale is on the length of the stencil of the underlying scheme, features of the boundary which are small enough to cause an issues in this regard are likely to be poorly resolved, so we suggest this is not in practice a severe limitation of the method.

In practice the main limitations we have found with the scheme presented here in fusion plasma devices are in problems where the interface motion is strongly sub-alfvenic but where magnetic fields shape the material interface, and in strongly shocked flows at low plasma beta (highly magnetized flows). 
In the former, the diffusivity of the base MHD scheme in the sub-alfvenic regime smooths out the magnetic field on a timescale faster than it can force the interface to change shape.
In the latter, the MHD scheme violates positivity of thermal energy.
Both of these limitations arise from the base MHD scheme adopted, are longstanding and well understood limitations, and arise before considering multiphase flows or flows with material interfaces.
As an explicit Godunov-type scheme the \cite{stone2009a} scheme is an  best suited to high-Mach number flows, but it is well known that schemes of this class struggle as the Mach number of the flow decreases \citep[e.g.][]{2010JCoPh.229..978D}.
To address low Mach number flow in purely hydrodynamic problems, schemes using either a low mach number modification of the Euler equations \citep{2006ApJ...637..922A,2011A&A...531A..86V} have been proposed, low Mach number Riemann solver schemes \citep{2015A&A...576A..50M,10.1007/s10915-017-0372-4}, and spectral difference schemes have shown promising results \citep{2025MNRAS.537.2387V}.
Recently developed all-Mach schemes appropriate for highly magnetized flows presented by \citet{2024JCoPh.51413229D} and \cite{2024arXiv240304517B}, if used as the base MHD scheme for a method like the one presented in this work, would provide a significant extension in capability into the regime of sub-alfvenic and highly magnetized flows.
The ghost fluid method \citep[GFM][]{fedkiw1999a} is known to produce spurious behavior when excessively large jumps in material properties are present at the interface \citep[see, e.g.][]{bigdelou2022a}. Thus, that limitation is inherited in our coupled Constrained Transport-GFM implementation but is not itself a limitation of our novel approach to solving the MHD equations.

\section*{Acknowledgments}
This work was supported by the U.S. Department of Energy through the Los Alamos National Laboratory. Los Alamos National Laboratory is operated by Triad National Security, LLC, for the National Nuclear Security Administration of U.S. Department of Energy (Contract No. 89233218CNA000001).
This work was supported by funding from the Government of Canada through its Strategic Innovation Fund (Agreement No.~811-811346). The funding source had no involvement in: the study design; the collection, analysis and interpretation of data; the writing of the report; the decision to submit the article for publication. 

\bibliography{references}
\bibliographystyle{aasjournal}

\end{document}